**Density Functional Theory screening of gas-treatment strategies for stabilization of high energy-density lithium metal anodes**

Stephan L. Koch,[a,b] Benjamin J. Morgan,[c,1] Stefano Passerini,[a,b] and Gilberto Teobaldi[c*]

a Helmholtz Institute Ulm (HIU), Helmholtzstrasse 11, 89081, Ulm, Germany

b Karlsruhe Institute of Technology (KIT), PO Box 3640, 76021 Karlsruhe, Germany

c Stephenson Institute for Renewable Energy, Department of Chemistry, University of Liverpool, L69 3BX Liverpool, United Kingdom

[1] Present address: Department of Chemistry, University of Bath, BA2 7AY Bath, United Kingdom

* Corresponding author: E-mail: g.teobaldi@liv.ac.uk (Phone: +44 151 795 8136)

**Abstract**

To explore the potential of molecular gas treatment of freshly cut lithium foils in non-electrolyte–based passivation of high-energy-density Li anodes, density functional theory (DFT) has been used to study the decomposition of molecular gases on metallic lithium surfaces. By combining DFT geometry optimization and Molecular Dynamics, the effects of atmospheric ($N_2$, $O_2$, $CO_2$) and hazardous ($F_2$, $SO_2$) gas decomposition on Li(bcc) (100), (110), and (111) surfaces on relative surface energies, work functions, and emerging electronic and elastic properties are investigated. The simulations suggest that exposure to different molecular gases can be used to induce and control reconstructions of the metal Li surface and substantial changes (up to over 1 eV) in the work function of the passivated system. Contrary to the other considered gases, which form metallic adlayers, $SO_2$ treatment emerges as the most effective in creating an insulating passivation layer for dosages ≤ 1 mono-layer. The substantial Li→adsorbate charge transfer and adlayer relaxation produce marked elastic stiffening of the interface, with the smallest change shown by nitrogen-treated adlayers.

# 1. Introduction

The increasing demand for stable, high energy density rechargeable batteries for long-range electric vehicles motivates the growing interest in developing alternative chemistry and cell strategies to replace existing Li-ion insertion-based technologies [1-2]. Driven by the substantial theoretical increase in energy density, great efforts are currently being devoted to the development of Li-air and Li-sulfur batteries [3-9], which depend on the (to date hypothetical) availability of stable, highly reversible, lithium metal anodes, capable of delivering a nearly ten-fold increase in theoretical capacity (3,860 mAh g$^{-1}$) over commercially used graphite anodes (372 mAh g$^{-1}$) [2].

The highly electropositive nature of Li (-3.04 V redox-potential vs. SHE [10]) is responsible for its extreme (reducing) reactivity towards exposed molecular media. Similarly to graphite anodes, initial decomposition of the electrolyte and the ensuing formation of a protective solid-electrolyte interphase (SEI), which should be Li$^+$-ion permeable yet mostly electronically insulating, is a beneficial process for the stabilization of Li anodes [1-9]. However, the repeated removal (stripping) and re-insertion (plating) of Li atoms beneath the SEI upon electrochemical cycling is known to cause cracks in the SEI and ensuing exposure of the electrolyte to metallic Li, leading to progressive electrolyte decomposition [11]. Another unresolved issue affecting the stabilization of Li metal anodes is the formation and growth, through cracks in the SEI, of highly reactive lithium metal protrusions (a.k.a. dendrites) during cycling. Lithium-dendrite growth eventually short-circuits the battery electrodes, which might cause an organic solvent electrolyte to ignite, leading to catastrophic failure of the battery [12-24]. Recent work indicates that besides cracking in the SEI during cycling, the presence of sub-surface impurities (nitrides and other compounds depending on the preparation/storage of the Li foil) at the Li anode can be critical for dendrite formation and growth [25].

Prevention of Li dendrites has so far focused on physical and chemical strategies to block their formation by controlling the SEI composition and morphology via use of pressure, SEI-stabilizing additives, ionic liquid-based electrolytes, as well as copolymer (solid) electrolytes and mixtures of liquid and polymeric electrolytes [14-15, 18-19, 23, 26-30]. Stabilization of Li anodes is made even more challenging by the simultaneous need for the electrolyte to decompose into electron-insulating SEI with sufficiently high diffusivity of $Li^+$ ions, allowing the flow of $Li^+$ ions to the cathode through a thermally and electrochemically stable electrolyte with a low boiling point. Although promising advances have been very recently shown to be possible via combination of carefully chosen liquid and polymer electrolytes with low reduction potential, high viscosity and large size anions [30], addition of halogenated salts (especially LiF) to the electrolyte [31], and hollow carbon nanosphere coating [32], stable cycling of Li anodes for several hundred cycles at room temperature at competitive (dis)charge rates (similar to those achieved with graphite) has not, to the best of our knowledge, been achieved yet.

The observed relationship between occurrence of cracks in the SEI and Li-dendrite formation suggests that creation of a *tough* (i.e. mechanical damage tolerant [33]) SEI should be beneficial in preventing dendrite formation. Recent research in damage tolerant natural and synthetic materials indicates that hierarchical multi-scale (nm to cm) structuring (extrinsic toughening [33]) of composite materials can be crucial for crack suppression [33-34]. However, the requirement of favorable $Li^+$ diffusivity through the SEI could be hardly meet by adoption of known extrinsic toughening strategies [33-34] leading to cm-thick SEI, which would exceed the thickness of commercial cells (both electrodes and electrolyte-soaked separator) by several orders of magnitude. These considerations make exploration of novel strategies towards formation of electrochemical and mechanical stable (nm-thick) SEI a necessity for viable stabilization of Li-anodes. To this end, the critical role of atomic relaxations for interface mechanical anomalies [35-38] and the expected substantial charge transfer involved in the SEI-formation call for atomistic insight into the structural and mechanical properties of Li anode SEI beyond available results from continuum models [24, 29].

Apart from, to the best of our knowledge, one exception where gas ($N_2$) pretreatment of metal Li was considered [39], the explored strategies for Li anode stabilization to date have invariably targeted formation of the SEI via decomposition of the cell electrolyte or components dissolved in it [14-15, 18-19, 23, 26-30]. Experimental work in the field has been complemented by a limited number of Density Functional Theory (DFT) studies of adatom energy and diffusion on vacuum-exposed [40] and implicitly solvated [41-42] Li surfaces, ionic liquid decomposition on defect-free Li(100) [43-44], force-field modelling of fractures in Li single crystal [45], and coarse-grained dynamic Monte Carlo studies of Li dendrites [46]. However, recent advances in the field indicate that major benefits can be achieved by pre-treating Li anodes before exposure to the cell electrolyte [39, 47]. In addition, the recently established link between Li subsurface impurities and dendrite formation [25] suggests that controlled deposition of impurities in metal Li substrates could be effective in preventing dendrite formation and growth.

Qualitatively, the ideal SEI or, as we start to explore here, an alternative passivation layer created by pre-treatment of the metal Li anode, should fulfill the following conditions: **i)** it should be electronically insulating in order to prevent electron transfer from the Li anode to the electrolyte. **ii)** It should be thick enough to suppress electron tunneling from the (biased) electrode to the electrolyte. To this end, we speculate that **iii)** the occurrence of a Li-SEI interface dipole opposing (zero-bias) electron-transfer from the passivated anode to the electrolyte may be beneficial. **iv)** The SEI should be tough [33] to adapt to the volume changes of the Li anode upon cycling (stripping during discharge and plating during charge) without cracking. Alternatively, **v)** a SEI capable of quickly self-healing [48-50] the cracks created during cycling may be also highly beneficial. **vi)** The SEI should allow good diffusivity of $Li^+$ ions. In this respect, nm-thick SEI (favoring Li-diffusion) may be preferable, provided the SEI is sufficiently thick to keep the Li-anode and the electrolyte electronically decoupled. Ideally, **vii)** it should be possible to tune a priori the $Li^+$ ion diffusivity of a given SEI to match the given cathode redox chemistry and (dis)charge rate, allowing for balanced battery assembly. **viii)** The SEI should be impermeable to (and insoluble in) the electrolyte solvent and other contaminants dissolved in it. Simultaneous fulfilment of all these conditions, and stability of the SEI over several (hundred to thousand) charge-discharge cycles is clearly a formidable challenge, which can be hardly met without a thorough understanding of the atomic-scale factors governing the SEI formation and evolution upon cycling.

The SEI formation via inherently out of equilibrium chemistry during the initial contact with the electrolyte and cycling of the Li anode, and the limited atomic scale control of the pristine Li surface present severe challenges to the characterization, thence understanding and eventual optimization, of the SEI formation in controlled and reproducible conditions. These considerations, encouraging results on the beneficial role of $N_2$ treatment of Li metal anode [39], and the observed dependence of Li anode stability on the inert atmosphere (e.g., dehydrated air vs. argon) in which commercial Li-foils are made [47, 51], make us wonder whether alternative gas–solid, equilibrium based, chemical strategies could be used to create a working (i.e. fulfilling conditions i-viii above) SEI or SEI-precursor layer on Li metal anodes, *before* contact with the liquid or polymeric electrolyte. To the best of this knowledge, this strategy has not been systematically studied, which motivates the present work.

Although the complexity and extension of the actual anode/SEI/electrolyte interfaces cycled at variable electrode bias is well beyond the current capabilities of standard Density Functional Theory (DFT) methods [52-53], previous success of DFT-based strategies in elucidating the reactivity of adsorbed molecules on Li metal surfaces [43-44, 54] makes the approach convenient for exploration of some of the benefits which might be achieved by gas treatment of pristine, freshly cut Li anodes. Furthermore, the adopted computational approach allows preliminary assessment of the actual benefits of using hazardous gases (i.e. $F_2$ and $SO_2$, vide infra) without taking unnecessary experimental risks.

To explore the opportunities offered by molecular-gas–passivation of pristine Li metal surfaces, here we investigate 0 K and room temperature decomposition of different molecular gases on metal Li surfaces for coverages in the 0.25-1 mono-layer (ML) range. Using DFT geometry optimization and Molecular Dynamics we investigate the effects of atmospheric ($N_2$, $O_2$, $CO_2$) and toxic ($F_2$, $SO_2$) gas decomposition on the relative energy of Li(bcc) (100), (110), and (111) surfaces, their reducing potential (approximated by the corresponding work function), and emerging electronic and elastic properties.

The presented results indicate that dosing of different gases can lead to passivation layers with profoundly different electronic and elastic properties. Depending on the dissociated gas, insulating or metallic adlayers, with surface elastic constants up to ten times stiffer or softer than the pristine Li-surfaces, can be obtained. We believe these results should be useful to inform future experimental efforts towards stabilization of high energy-density Li-metal anodes via gas pre-treatment of the Li electrodes.

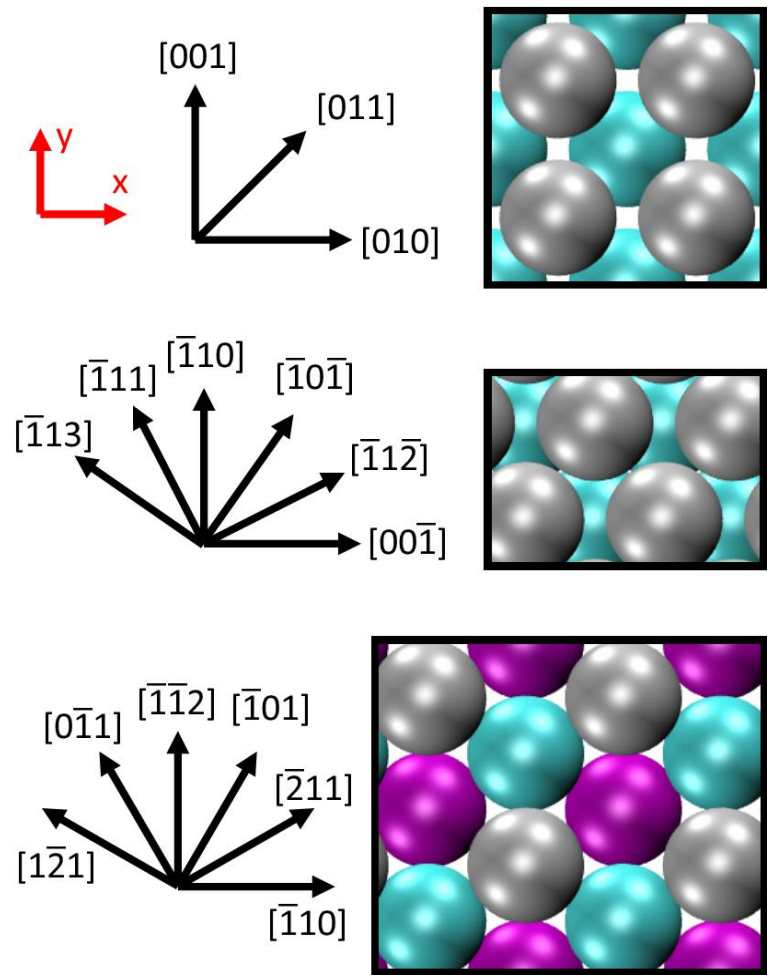

**Figure 1.** Top-view of the adopted Li(100)-2×2 (top) , Li(110)-2×1 (middle) and Li(111)-2×1 (bottom) orthorhombic slabs, together with the surface high symmetry directions. The black square (rectangle) marks the in-plane periodicity of the simulation cells. Topmost Li-layer: grey. 2$^{nd}$ topmost Li-layer: cyan. 3$^{rd}$ topmost Li-layer: purple.

## 2. Methods

DFT simulations were performed via the Projected Augmented Wave (PAW) method as implemented in the VASP program [55]. In all cases, the PBE exchange-correlation (XC) functional [56], a 400 eV plane wave energy cutoff, and 0.2 eV Gaussian smearing were used. At least 15 Å vacuum separation between periodic replicas of the slab models and dipole corrections were used for all the simulations. To prevent introduction of artificial dipoles perpendicular to the surfaces, molecules were adsorbed on both sides of the slabs. Geometry optimizations were performed without any atomic or symmetry constraints, with a force-convergence threshold of 0.05 eV Å$^{-1}$ via the RMM-DIIS quasi-newton algorithm [57]. The slab models of the Li(100), Li(110) and Li(111) surfaces were constructed using the DFT-optimized lattice constant for Li(bcc), 3.466 Å. In all cases, orthorhombic [Li(110) and Li(111)] and tetragonal [Li(100)] cells were used. The slab thickness [11-Li layers for Li(100) and Li(110), 21 layers for Li(111)] was chosen as the thinnest possible to yield surface energies and work functions converged to within 10 meV (see supplementary material). To allow for adsorption of the considered molecular gases, while containing the computational cost of the simulations, Li(100)-2×2, Li(110)-2×1 and Li(111)-2×1 supercells (of the orthorhombic and tetragonal unit cell) were modeled (Figure 1) with 7×7×1 [Li(100)], 7×10×1 [Li(110)] and 5×6×1 [Li(111)] grids of symmetry-irreducible k-points. The grids were chosen to maintain the same k-point spacing ($\leq 0.003$ Å$^{-1}$), which was checked to yield energies converged to within less than 1 meV/atom for bulk Li(bcc) (13 symmetry irreducible k-point grid). This choice produced (100), (110) and (111) slabs with four top Li adsorption sites per exposed surface (Figure 1).

Given their use for equilibration-only purposes, canonical (*NVT*) Molecular Dynamics (MD) simulations were run using the Verlet integration algorithm [58] and the Berendsen thermostat [59] as implemented in VASP. In all cases the time step was 1.5 fs. Geometry optimization and MD runs for all the molecularly decorated slabs were carried out allowing unconstrained spin-polarization in the system.

In analogy with previous studies of chemical bonding of organics at metallic surfaces [60-61], and as also discussed in [41], we found that use of van der Waals corrections (according to Grimme's parameterization [62]) negligibly affected the optimized geometry (< 0.01 Å changes in the adsorption length of the energetically favored systems).

Vibrational frequencies were calculated via symmetric finite displacements (± 0.05 Å) following further optimization (to within 0.01 eV Å$^{-1}$ force-tolerance) of the selected systems with an increased plane wave energy cutoff of 600 eV. Elastic tensors accounted for ion-relaxation following the procedures described in [63] and [64], as implemented in VASP. Based on the numerical (non-zero) value of the elastic constants bound to be zero owing to the symmetry [65] (orthorhombic or tetragonal) of the cells, the error of the procedure is < 0.1 GPa for the bare Li slabs.

Bader charge analyses [66] were carried out on the basis of the total charge density i.e. accounting for both the electronic and ionic core charges.

Slab formation energies ($E_{form}$) were calculated as:

$$E_{form} = E_{slab} - N_{Li}E_{Li-bulk} - N_{mol}E_{mol} \qquad (1)$$

where $N_{mol}$ is the number of gas molecules initially present in the system and $E_{mol}$ is the energy of one molecule optimized in vacuo.

Work functions (*W*) were calculated from the difference between the vacuum-electrostatic plateau ($E_v$) and the computed Fermi energy ($E_F$):

$$W = E_v - E_F \qquad (2)$$

## 3. Results and discussion

### 3.1. Choice of Li-surfaces and molecular gases

To explore the effects of molecular gas adsorption on Li(bcc) substrates, we considered three Li(bcc) surfaces with different surface-energy [67]. Specifically, we focused on the lowest surface-energy cuts [Li(100)], Li(110) and Li(111). According to recent plane wave PAW-PBE DFT simulations [67], Li(110) and Li(111) have a surface-energy comparable with and higher than other low-symmetry [(120), (133), (311)]] terminations, respectively. In line with this latter study, and in agreement with earlier (atomic basis set) DFT-studies [68], the computed surface-energy for Li(100) is lower than for Li(110), which in turn has a lower surface-energy than Li(111) (see supplementary material).

In this study we modelled the adsorption of atmospheric gases ($N_2$, $O_2$, $CO_2$) and lone-pair rich molecular gases ($F_2$ and $SO_2$). This choice was driven by fact that $N_2$ (78.08 %), $O_2$ (20.95 %) and $CO_2$ (0.04 %), together with inert Ar-gas (0.93 %), are the main gases present in dry-air [69] in which commercial Li-foils and Li-anodes are routinely prepared. The study of $F_2$ and $SO_2$ adsorption was prompted by their hazardous nature (thence experimental reluctance for their handling), experimentally observed improvements in Li anode SEI upon addiction of halide (fluoride) salts to the electrolyte [31], and speculation that dissociative adsorption of lone-pair rich systems, potentially leading to a lone-pair rich passivation layer, benefit $Li^+$ coordination and diffusivity across the layer.

### 3.2. Optimized molecular adlayers

For all the three considered crystallographic cuts [Li(100), Li(110), Li(111)] and gases, initial geometries were prepared for different coverage in the 0.25–1 ML range placing the undissociated adsorbate on different surface adsorption sites (Figure 1) at distance of at least 1.8 Å from the topmost Li-atoms. For all considered gases, several different initial molecular orientations were explored, with more than 60 adsorption geometries being screened for each gas. Details on the initial geometry set up and energy screening after geometry optimization can be found in the supplementary material.

Tables S1-S6 in the supplementary material list all the considered initial geometries, together with the computed slab formation energy ($E_{form}$) after geometry optimization. In all cases we model strongly exothermic ($E_{form} < 0$) reaction between the molecular gases and the Li-surfaces, accompanied by significant rearrangement of the topmost Li-layers. Unsurprisingly, given their known large electronegativity and oxidizing chemistry [10], $F_2$, $O_2$ and $SO_2$ yield the lowest $E_{form}$ when reacted with Li-slabs. Reaction with $N_2$ and $CO_2$ is computed to be substantially less exothermic (less negative $E_{form}$). Figure 2 summarizes the computed lowest $E_{form}$ for each molecular gas on Li(100), Li(110) and Li(111). The atomic structure of the lowest $E_{form}$ systems for each gas is shown in Figure 3. The SI reports atomistic models of the lowest $E_{form}$ system for all the considered Li surfaces.

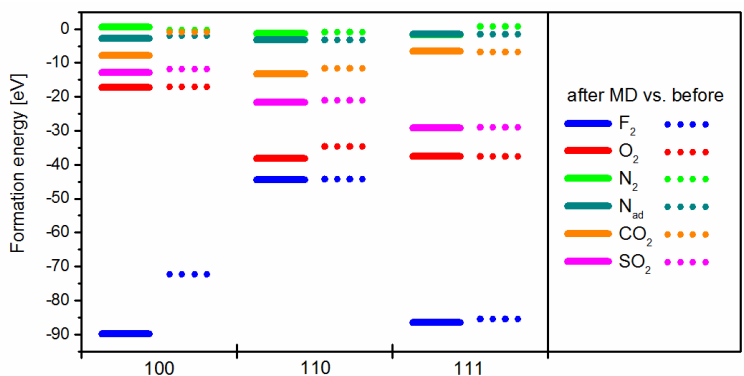

**Figure 2.** Lowest slab formation energies ($E_{form}$, eV) following optimization (dotted lines) of each of the considered molecular gases on Li(100), Li(110), and Li(111). The results for the system optimized after *NVT* MD at 300 K equilibration are shown as continuous lines.

Molecular adsorption on the Li-slab strongly affects the surface energy ranking, with the (100) termination, favored in vacuo, being replaced by the (111) ($F_2$, $O_2$, $SO_2$) or (110) ($N_2$, $CO_2$) slabs as the energetically favored substrate. These results suggest that, at least on an energy basis, molecular adsorption induced reconstruction of Li-surfaces could be viable, thence in principle engineered by controlling the atmosphere during preparation of Li anodes.

Alkali metal redox chemistry with aqueous and other reducible media is known to be extremely vigorous and potentially explosive depending on the mixing of the preliminary products [70], which is reflected in the computed very negative (< 10 eV or, equivalently, < 2.9 eV / adsorbate) $E_{form}$ following dissociative adsorption of $F_2$, $O_2$, $SO_2$. The substantial energy released upon dissociative adsorption of $F_2$, $O_2$, $SO_2$ may be effective in promoting adsorbate induced Li-surface reconstruction of freshly cut Li-surfaces. Alternatively, co-dosing of small amount of $F_2$, $O_2$, $SO_2$ during initial gas treatment of freshly cut Li anodes, and the ensuing energy release, could be used to activate and/or alter the surface chemistry with other, less reactive, molecular gases.

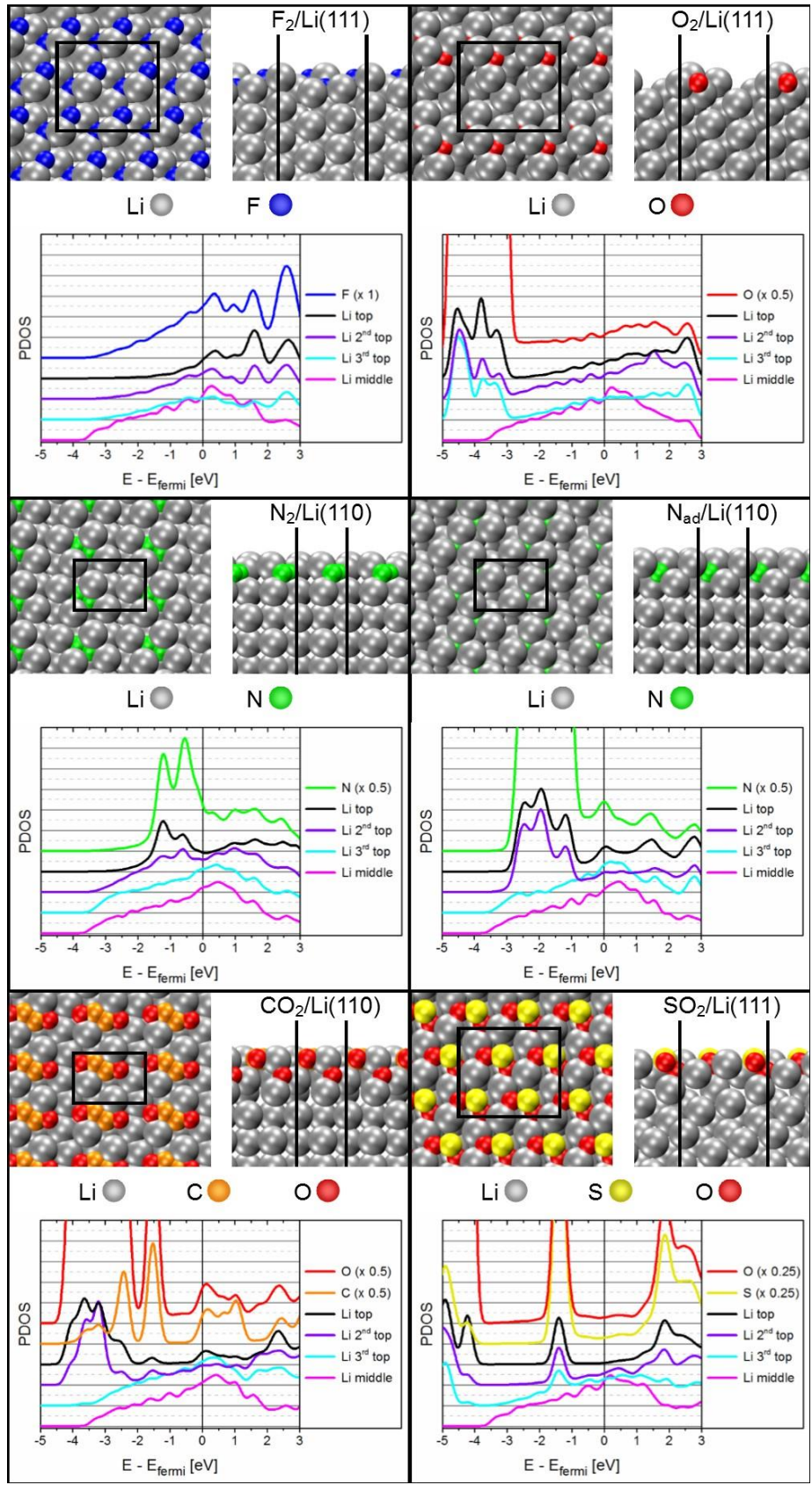

**Figure 3.** Optimized geometry and layer-resolved atom-projected Density of States (PDOS) for the lowest $E_{form}$ systems (Figure 2) of each considered molecular gas.

Unsurprisingly, Bader charge analysis for the lowest $E_{form}$ systems (Figure 4) reveals substantial charge transfer (> 1.4 e/molecule) from the Li-slabs to the adsorbed molecules, the only exception being the weakly bound $N_2$/Li(100) system ($E_{form}$ = -0.24 eV). The charge transferred from the Li-slab to the chemically bound systems is largest for the $O_2$/Li(100) adlayer (4.25 e / molecule) and smallest for $CO_2$/Li(100) (1.45 e / molecule). The computed trends do not follow the experimental values of electron affinity (*EA*) for the considered gases [$EA(CO_2)$ = 3.225 eV < $EA(F_2)$ = 3.01 eV < $EA(SO_2)$ = 1.107 eV < $EA(O_2)$ = 0.451 eV [71-72]], indicating a non-negligible role of the gas-induced structural rearrangement for total amount of charge transferred from the Li-slab. Further evidence of the intricate interplay between charge transfer and structural rearrangement for the energy of the molecularly decorated slabs is found in Figure 4. The computed $E_{form}$ for different molecular systems (and for the same gas adsorbed on different Li slabs) does not directly correlate with the amount of charge transferred to the adsorbed molecules, the only exception being $CO_2$, for which the calculated $E_{form}$ decreases with increasing charge transfer.

Overall, these results clearly indicate that designing molecular-gas treatment of Li-slabs towards engineering of pristine passivation layers based on the experimental (or computed) EA of the molecular reactants could be highly misleading. Direct simulation of the reaction products turns out to be necessary for rational development of experimental gas-treatment strategies towards passivation of Li-substrates.

Inspection of the optimized geometry for the lowest $E_{form}$ systems reveals dissociation for $F_2$, $O_2$, and $SO_2$, molecular condensation for $CO_2$, (formation of an adsorbed acetylenediolate, $C_2O_2$, species) and subsurface intercalation for $N_2$, $N_{ad}$, and $SO_2$ (O-atoms). The supplementary material contains further analysis of the optimized geometry. It is worth to recall that these results have been obtained following geometry optimization, which indirectly points to the existence of barrier-less reaction and intercalation channels for the considered gases on Li surfaces (from the adopted initial geometries). Consistent with recent DFT results on solvated alkali metal (Na) clusters [70], we find that atomic relaxation is not needed to trigger initial charge transfer at the immediate Li/adsorbate interface. This electron transfer strongly alters the potential energy surface governing the molecular and interface relaxation, leading to barrier-less reaction for all the considered adsorbates.

Whereas formation of oxide dissociation products following $O_2$ adsorption is consistent with available XPS results for $O_2$-treatment of metal Li films [73], the occurrence of an acetylenediolate $C_2O_2$ product (and oxide subsurface intercalation) from $CO_2$ dissociation does not match experimental XPS suggestions of oxalate ($C_2O_4$) intermediate formation on Li from the reaction of $CO_2$ with metal Li at 120 K [73-78]. While these deviations could be caused by biases in the simulations due to the limited size of the simulation cells and neglect of surface defects as well as temperature effects, we note that in [73] the Li-substrate was characterized (at 120-350 K) after substantially larger (30 Langmuir) molecular exposure then considered here, which may explain the observed differences. Although the simulations suggest that formation of adlayers with isolated N-adatoms is energetically favored over $N_2$ subsurface intercalation ($N_{ad}$, Figs 2,3), it is interesting to note that, in spite of the substantial charge transfer (Fig. 4), intercalation of (markedly elongated to 1.34 Å, supplementary material) $N_2$ molecules turns out to be favored over $N_2$ dissociation (at 0 K) on defect–free substrate. Based on the experimentally known occurrence of nitride ($Li_3N$) contamination in $N_2$-exposed Li-foils [25], we speculate that $N_2$ dissociation may be triggered at surface defects (neglected in our models).

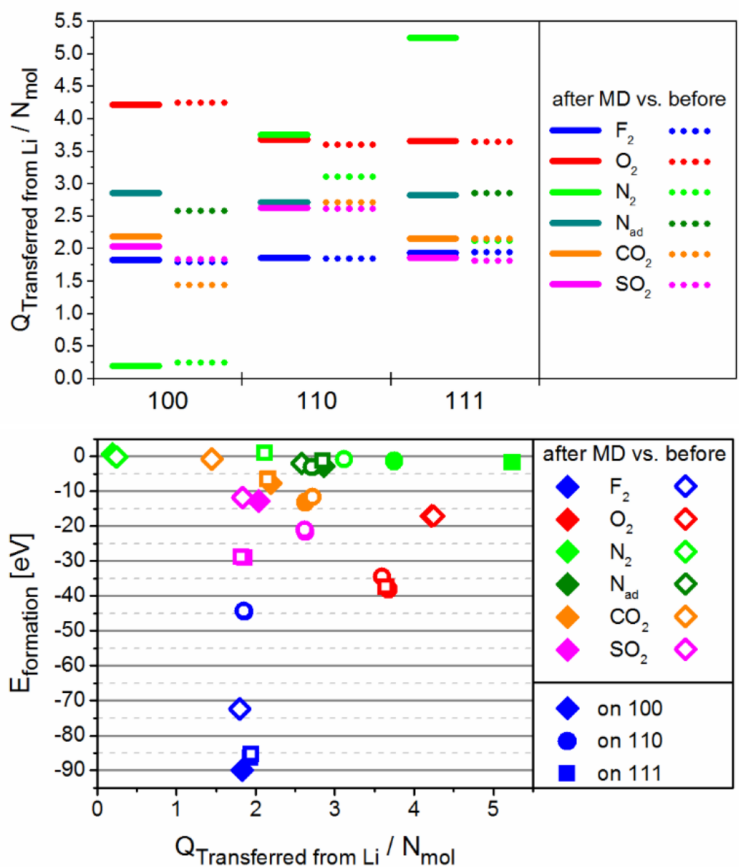

**Figure 4.** Top: Bader electronic charge (per molecule) transferred from the Li-slabs. Bottom: Slab formation energies ($E_{form}$, eV) as a function of the Bader electronic charge (per molecule) transferred from the Li-slabs.

Electronic decoupling between the metallic electrode and the electrolyte is a key requirement for any Li anode passivation strategies. Any viable SEI or passivation layer needs to be insulating *i.e.* have a band gap at the anode $E_F$ suppressing electron transfer, thence redox chemistry, between the (biased) anode and the electrolyte. To investigate whether the lowest $E_{form}$ adlayers on Li-surfaces are effective in creating an insulating protective cap, we analyze the Total Density of States (DOS) and atom-projected DOS (PDOS) on a layer resolved basis.

With the exception of $SO_2$, layer-resolved analysis of the PDOS for the lowest $E_{form}$ systems (Figure 3 and supplementary material) gives evidence of metallization for all the adlayers. Thus, dissociation of 0.25-1 ML $F_2$, $N_2$, $O_2$, or $CO_2$ turns out to be insufficient to create an insulating thin-film on Li(100), Li(110), and Li(111). The occurrence of metallic, therefore arguably conducting, Li-adsorbate reconstructions exposed to the medium suggests that larger molecular dosages (> 1 ML) are needed to grow thicker, expectedly insulating, passivation layers capable of electronically decoupling the anode and electrolyte. The $SO_2$ case stands apart from the others since the adsorbate and topmost Li layers reveal a noticeably suppressed PDOS at $E_F$. These results suggest that low-dosage $SO_2$ treatment should be more effective than low-dosage $F_2$, $N_2$, $O_2$ and $CO_2$ exposure in creating ultra-thin insulating passivation layers, which may be beneficial for extremely fast $Li^+$ diffusion.

The characteristically strong reductive chemistry of metal Li (and other alkali metals) is intimately related to its high $E_F$ value, or equivalently, low work-function ($W$ = 2.9 eV for polycrystalline metallic Li [79]) in comparison to more inert transition metals (> 4.5 eV [67, 79]). Accordingly, increase of metal Li $W$ by molecular passivation, resulting in an energetically more costly electron extraction, hence lower $E_F$ and expectedly lower reducing reactivity may be a rewarding strategy towards stabilization of Li-anodes. Figure 5 compares the calculated $W$ for clean Li-slabs and the lowest $E_{form}$ slabs for each gas. With the exception of $SO_2$/Li(111) and $SO_2$/Li(110), the calculated $W$ for the $E_{form}$–favored systems is up to 1.2 eV smaller than for the pristine clean surfaces. These computed lower $W$ values indicate more favorable electron extraction from the slab, corresponding to potential enhanced reduction of the electrolyte. It transpires, therefore, that low-dosage molecular treatment of Li-slabs as considered here would *increase*, rather than quench, the reducing reactivity of Li-slab. The only exceptions are $SO_2$/Li(111) and $SO_2$/Li(110) for which we compute a noticeable increases of increase in $W$ (+0.3 eV and +0.8 eV, respectively), which in turn suggest decreased reducing reactivity of the passivated slab.

It is interesting to note that for higher $E_{form}$ structures, which are therefore predicted to be less frequently observed, such as the lowest $E_{form}$ $CO_2$/(100) system (Figure 2), the calculated $W$ increases by up to 2 eV with respect to the value for clean Li(100). This indicates that just by adsorption of ≤ 1 ML of different gases, and as a result of the different adlayer geometries and Li-adsorbate charge transfer, an engineered increase of metal Li *W* by more than 1.5 eV could be in principle possible. Further work is in progress to investigate the evolution of the computed changes in metal Li *W* for larger dosages of molecular gases. These results will be reported elsewhere.

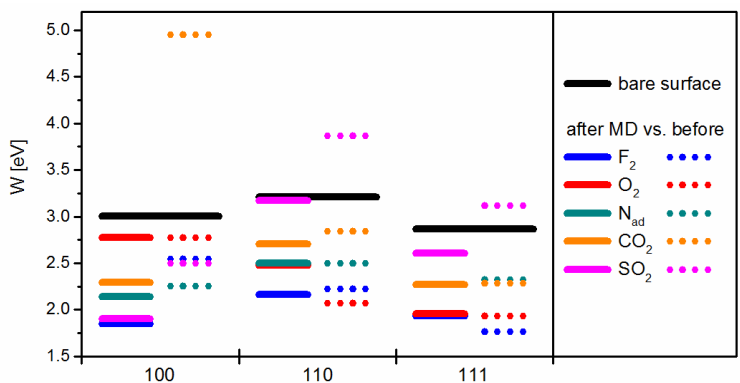

**Figure 5.** Calculated work function (*W*, eV) for the lowest $E_{form}$ optimized systems (dotted lines) on Li(100), Li(110), and Li(111). The results for the system re-optimized after MD 300 K equilibrations are shown as continuous lines. The result for the clean Li(100), Li(110) and Li(111) surfaces are shown as black continuous lines.

### 3.3. Optimized molecular adlayers following *NVT* MD equilibration

To investigate the occurrence of artefacts in the structural screening by geometry optimization of structures prepared starting from undissociated gas molecules, the lowest $E_{form}$ system for each considered gas and Li-slab were subject to a short (> 1.5 ps) *NVT* MD equilibration at 300 K (supplementary material) followed by geometry optimization of the final MD snapshots. The optimized structures were then subject to structural and electronic characterization (Figure 2, 4-6). We stress that rather than statistically robust insight into the real-time dynamics of the optimized adlayer (which would require larger simulation cells, longer MD trajectories, and rigorous canonical ensemble sampling [80]), the main target of this study was to use *NVT* MD equilibration, followed by structural relaxation, to identify lower $E_{form}$ minima potentially missed by the initial screening. This is a simple form of simulated annealing that permits the system to escape from high energy metastable configurations (local minima on the global potential energy surface).

The largest deviations in structure and $E_{form}$ (> 6 eV) take place for $F_2$/Li(100) and $CO_2$/Li(100), with smaller yet quantitatively significant (> 0.1 eV) changes for all the other systems apart from $O_2$/Li(111 and $N_{ad}$/Li(110) (Figure 2; see also supplementary material). As a result of these deviations, apart from $SO_2$ [$CO_2$ and $N_{ad}$], for which adsorption on Li(111) [Li(110)] remains energetically favored, the relative energy of the reconstructed systems deviates from what obtained after the initial geometry optimizations. Specifically, $F_2$/Li(100) and $O_2$/Li(110) are substituted to $F_2$/Li(111) and $O_2$/Li(111), respectively, as the lowest $E_{form}$ systems. Furthermore, Li(111) is computed to be the surface yielding the lowest $E_{form}$ minima for $N_2$ adsorption, in contrast to the results of the original screening, which suggest $N_2$/Li(110) to be energetically favored.

The changes in $E_{form}$ (Figure 2) reflect changes in the charge transferred from the Li-slab (Figure 4) as a result of the different adlayer rearrangement (Figures 4 and 6). These changes, however, do not affect the conclusion that, apart from $CO_2$, the computed $E_{form}$ for adsorption of $F_2$, $O_2$, $N_2$ and $SO_2$ is not directly correlated with the charge transferred from the Li-slab.
Inspecting the optimized geometry for this second set of lowest $E_{form}$ systems (Figure 6) confirms dissociative adsorption for $F_2$, $O_2$, and $SO_2$, as well as molecular condensation for $CO_2$ with formation of an adsorbed acetylenediolate, $C_2O_2$ species and subsurface O-atom intercalation. Subsurface intercalation is consistently predicted (observed? predicted?) also for the energetically favored adlayer of $N_2$ and $N_{ad}$.

The computed work function (*W*) for these new lower $E_{form}$ structures results is invariably smaller (up to more than 1 eV) values with respect to the pristine metal Li slabs (Figure 5) for all considered gases, $SO_2$ and $CO_2$ included. These results confirm that that low-dosage molecular adsorption (0.25–1 ML), as considered here, is not effective in creating an interface dipole capable of lowering the slab $E_F$, increasing its *W*, thence quenching the reducing activity of the metal Li substrates. To this end, larger dosage (> 1 ML) may be effective. Work in this respect is in progress and will be the subject of a forthcoming contribution.

Layer-resolved analysis of the PDOS for the lower $E_{form}$ minima after *NVT* MD equilibration confirms that adsorption of $SO_2$ on Li(111), leading to S-, O- and topmost Li-PDOS suppression at $E_F$ is still more effective then adsorption of $F_2$, $O_2$, $N_2$ and $CO_2$ in creation of a nearly insulating passivation layer capable of electronically decoupling the Li-subsurface from the exposed medium.

Overall, the modelled changes in structure, $E_{form}$ and *W* for the optimized system before and after *NVT* MD equilibration suggest that extra care should be taken when modelling molecular adsorption on Li slabs in the absence of experimental structural input. For the considered systems, structural screening via geometry optimization alone is shown to be clearly not sufficient.

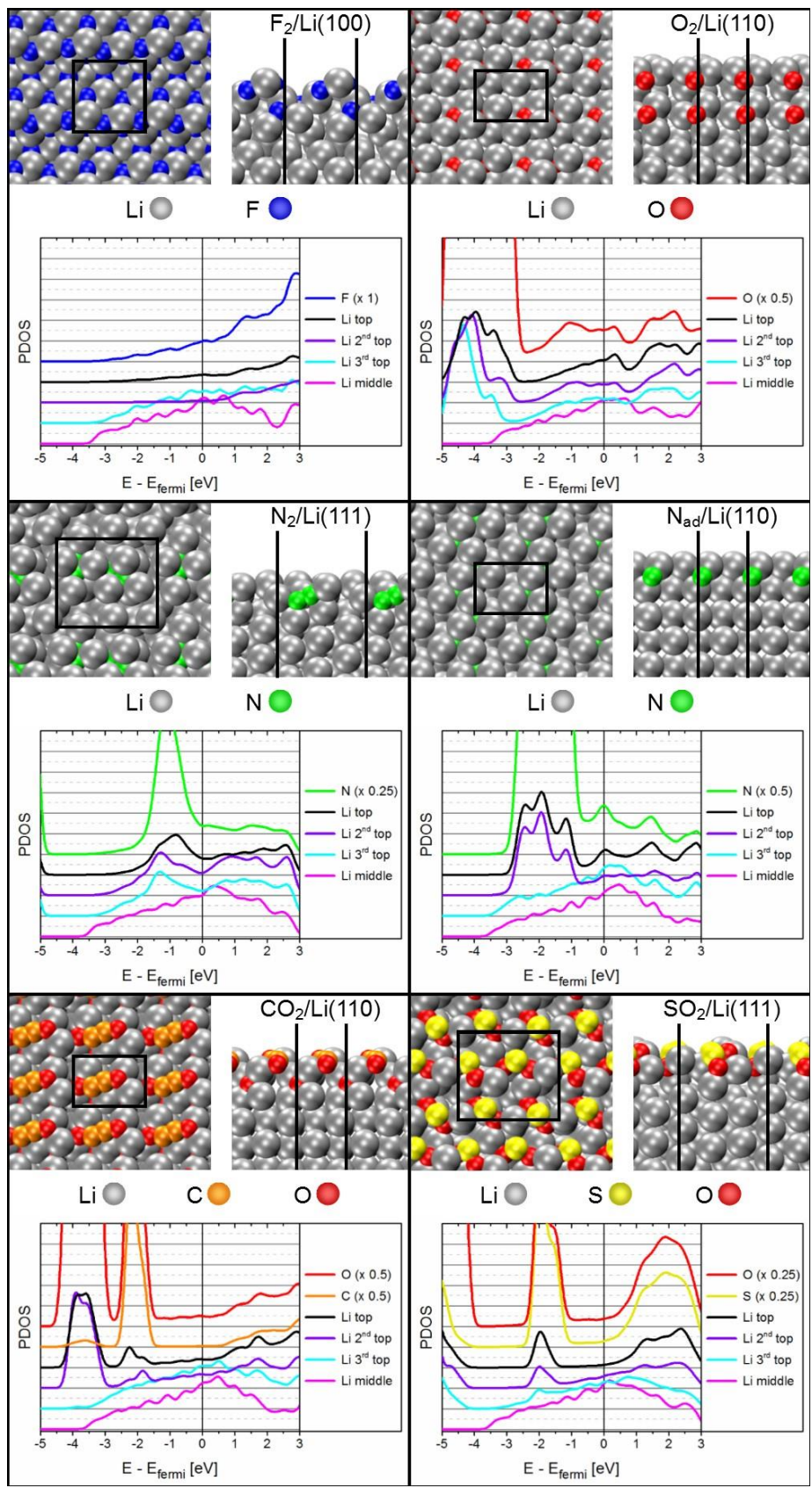

**Figure 6.** Optimized geometry and layer-resolved atom-projected Density of States (PDOS) for the lowest $E_{form}$ systems (Figure 2) of each considered molecular gas optimized after *NVT* MD equilibration.

## 3.4. Vibrational and elastic properties of the passivated Li slabs

We now analyze the effects of the molecular dissociation on the elastic properties of Li slabs. Although the limits of mechanical interface stabilities for Li-surface contacted to a polymer electrolyte have been previously studied via continuous elastic theory [29], the substantial rearrangements and charge transfer at the Li-adsorbate interface (Figures 3, 4, and 6), and ensuing likely change of the interface elasticity from the bulk counterparts (an aspect neglected in the continuous treatment [29]), calls for atomistic investigations of the elasticity of the passivation layers. Surface and interface atomic relaxations are known to strongly affect the elastic properties of layered or nanostructured materials as well as grain-boundaries [35-38]. The observed correlation between Li dendrite growth and crack formation due to mechanical stress of the SEI [11-23, 25-30] motivates our interest in the dependence of the Li-adlayer elastic properties on the composition of the dissociated gas, amount of interface re-organization, and Li-adsorbate charge transfer. We speculate that elastically compliant passivation layers with small elastic constants, leading to reduced propensity for irreversible plastic deformation and crack formation, should be beneficial for stabilization of Li-anodes. How to maximize the elastic compliance of Li-passivation layers has, to best of our knowledge, received no attention in the atomistic simulation literature, which prompts this section.

### 3.4.1. Harmonic Vibrational Frequencies

As a first approximation to the elastic properties of the passivated Li-slab, we computed the harmonic vibrational frequencies for the molecularly decorated slabs. To a first very qualitative and exploratory approximation (to be rigorously tested in the next section), one could suspect that the introduction of hard (high frequency) adlayer vibrations may lead to a stiffer (i.e. less elastically compliant) Li-adsorbate interface.

For this analysis we consider the bare Li(100), (110) and (111) surfaces together with the lowest $E_{form}$ system for each considered molecular gas. The computed (Γ-point) vibrational frequencies are shown (as wavenumbers) in Figure 7. The largest wavenumber vibrations for the bare Li slabs are below 340 cm$^{-1}$. In all cases, we compute a noticeable vibrational hardening, with at least 1.8-fold increase of the highest energy vibrational modes for F$_2$/Li(100). While the computed vibrational hardening for O$_2$, N$_{ad}$ and SO$_2$ is somehow larger ($\times$ 1.9-2.5 increase), the acetylenediolate C–O and C–C stretchings for CO$_2$/Li(110) (Figure 6) leads to a much larger ($\times$ 6.5) increase. Although the adopted approximated DFT functional (PBE [56]) is well known to generally yield underestimated vibrational frequencies [81], we expect the relative molecule-induced vibrational hardening to be qualitatively correct.

Using the vibrational hardness of the adlayer as an approximated measure of its elastic compliance, it is tempting to link the smallest computed vibrational hardening of F$_2$/(100) with recent reports on the beneficial role that addition of LiF to the electrolyte, and ensuing formation of LiF-rich SEI, plays in the stabilization of Li-anodes [31]. Along the same line, and owing to the ×6.5 increase in the adlayer vibrations and expected reduction in the elastic compliance of the passivated Li-slab, the simulations suggest that initial low dosage CO$_2$ treatment of Li substrates could be a very detrimental choice, which should be accordingly avoided. In the next section we test these results against explicit evaluation of the slab elastic constants.

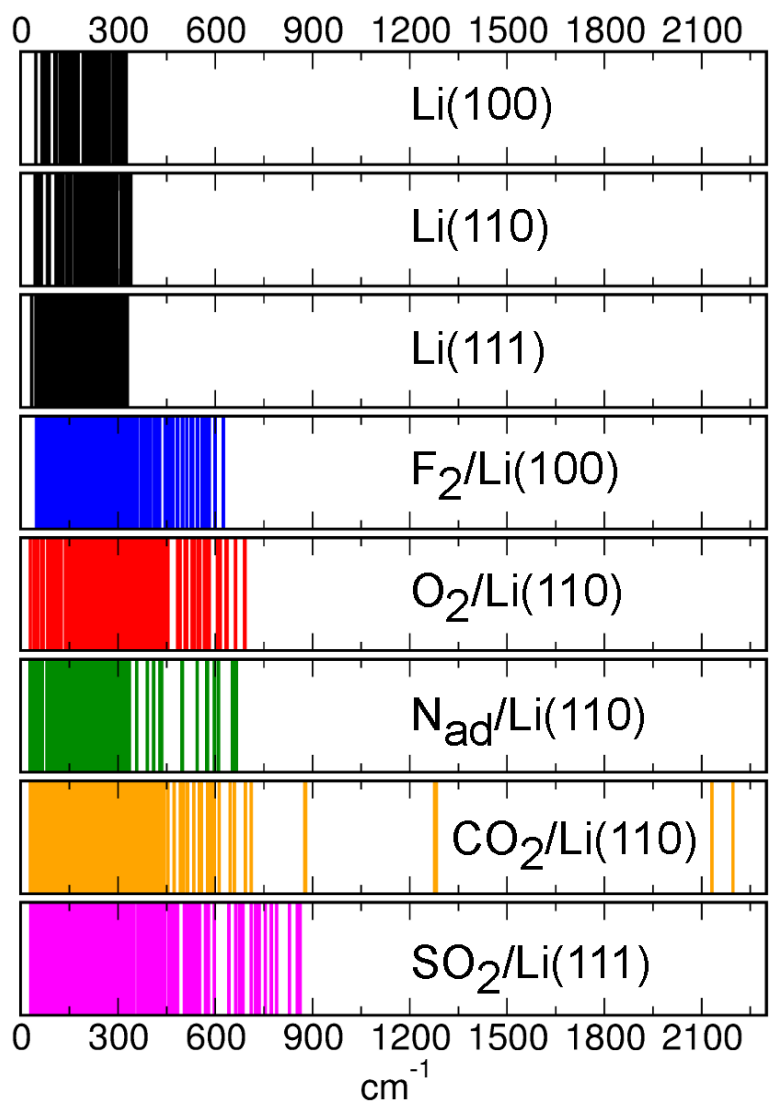

**Figure 7.** Computed vibrational (Γ-point) wavenumbers (cm$^{-1}$) for the bare Li slabs and the lowest $E_{form}$ systems (Figures 2 and 6) of each considered molecular gas after *NVT* MD equilibration.

**3.4.2. Elastic constants**

Investigation of the elastic properties of the passivated Li slab is extended by explicit evaluation of the surface elastic constants for the lowest $E_{form}$ systems. We are particularly interested in the effects which adsorption of different molecular gases cause on the elastic compliance of the passivated slabs. From a specialist perspective, and aiming at reducing the computational cost of further molecular screening, we are also interested in testing the (un)suitability of using the adlayer vibrational hardness as an approximated measure of its elastic compliance.

According to linear elasticity theory, the elastic constants of a system (collected in the stiffness tensor **C**) govern the proportionality between the stress (tensor, **σ**) generated in an isotropic material and the applied strain (tensor, **ε**) [65]:

$$\boldsymbol{\sigma} = \boldsymbol{C}\boldsymbol{\varepsilon} \tag{3}$$

From Eq. (3) it follows that the larger (and positive) the component of the elastic tensor, the larger the stress generated for the same experienced strain. Accordingly, elastically compliant interfaces with small **C**-components would be clearly desirable to minimize the stress generated by Li$^+$ diffusion across the passivation layer and the ensuing strain.

For the adopted orthorhombic Li(110) and Li(111) slab geometry (Figure 1, 2mm plane point group), there exist five independent in-plane elastic constants $C_{ijkl}$ ($C_{xxxx}$, $C_{yyyy}$, $C_{xyxy}$, $C_{xxyy}$, and $C_{yyxx}$ in extended Cartesian notation [37-38, 65]). The number of independent in-plane elastic constants reduces to three ($C_{xxxx} = C_{yyyy}$, $C_{xxyy} = C_{yyxx}$, $C_{xyxy}$ [37-38, 65]) for the tetragonal Li(100) slab (4 mm plane point group [82]). Figure 8 displays the independent in-plane elastic constants for the bare Li slabs and the lowest $E_{form}$ system for each considered gas.

In stark contrast to the results of the harmonic vibrations analysis, which suggests the smallest vibrational hardening for F$_2$/Li(100) (Figure 7), the computed elastic constants reveal that F$_2$ adsorption actually causes the largest elastic stiffening, with a nearly ten-fold increase of $C_{xxxx}$ (= $C_{yyyy}$) and $C_{xyxy}$. Substantial and strongly anisotropic elastic stiffening is modelled also for SO$_2$/Li(111), with increase in the elastic constants ranging from 20% ($C_{yyyy}$) to 65% ($C_{xxyy}$ and $C_{yyxx}$). The calculated changes in the elastic properties of O$_2$/Li(110) and CO$_2$/Li(110) are more complicated, with softening (reduction) of $C_{xxxx}$, minimal changes for $C_{yyyy}$ and $C_{xyxy}$ and opposite changes in $C_{xxyy}$ and $C_{yyxx}$ (CO$_2$: 30% stiffening, O$_2$: 70% softening). Overall, the N$_{ad}$/Li(110) elastic constants shows the smallest deviations (< 25%) from the bare Li(110) slab. Thus, in spite of its metallicity (Figure 6), and substantial charge transfer from the Li slab (2.71 e/molecule, Figure 4) creation of a thin adlayer with subsurface N-atoms [energetically favored on Li(110)] could be beneficial thanks to the formation an interface of elasticity comparable to the pristine Li substrate.

As shown in Figure 8, with the exception of $F_2$/Li(100), the computed elastic stiffening (increase in $C_{ijkl}$) and softening (decrease of $C_{ijkl}$) do not correlate directly with the computed slab formation energy ($E_{form}$), which rules out also approximation to the adlayer elastic stiffening on the basis of computed formation energy or charge transfer (Figure 4).

Although it is tempting to link the smallest computed elastic stiffening of $N_{ad}$/Li(110) with the measured increased cycling efficiency of Li metal anodes passivated with $N_2$ gas-solid treatment [39], one must be cautious: The elastic constants of the composite Li-passivation layer will inevitably evolve with increased thickness (larger molecular gas dosage) of the adlayer. Accordingly, the results obtained for 0.25-1 ML coverage should not be taken as representative of a nm-thick SEI as present in reality [39]. Further work will focus on the study of the dependence of the passivation-layer elasticity on its thickness and structure.

To summarize this section, analysis of the elastic constants for the molecularly decorated slabs, strengthen previous conclusions on the complex role of atomic relaxations (and, based on this work, charge transfer) for the elastic properties of (Li-adsorbate) interfaces [35-38]. Perhaps unsurprisingly, the adlayer vibrational hardness, formation energy and Li→adsorbate charge transfer are found not to directly correlate with the adsorbate-induced elastic stiffening/softening of the slabs, suggesting that explicit evaluation of the elastic constants of a given passivation layer cannot be avoided to quantify its elastic compliance.

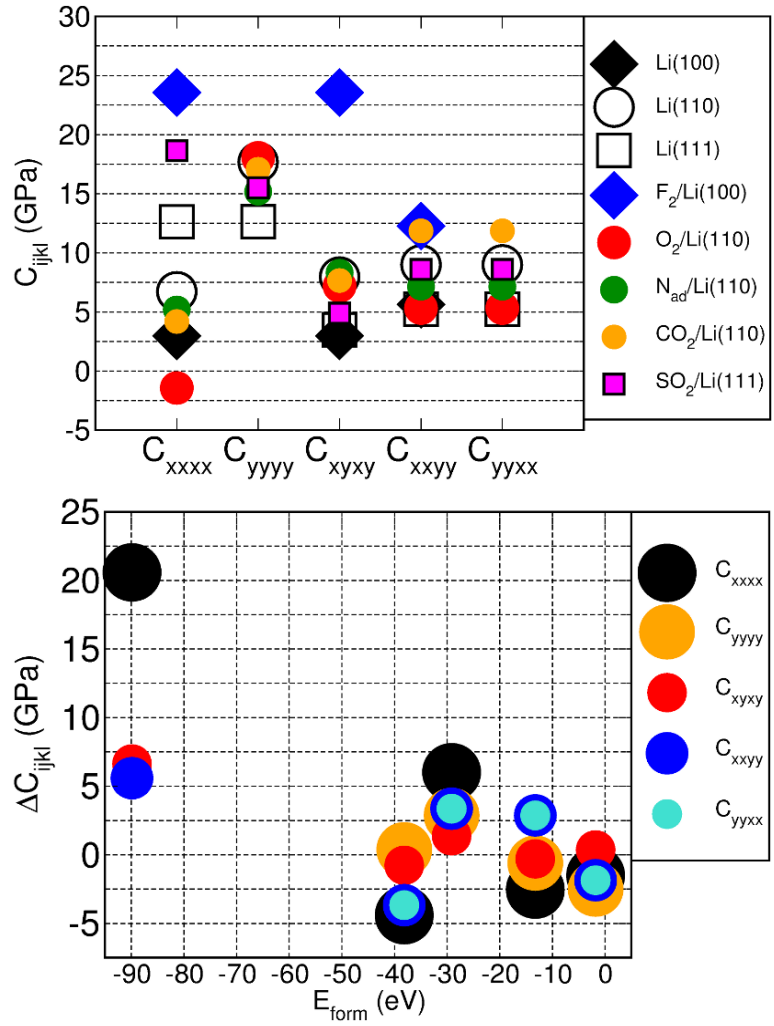

**Figure 8.** Top: Computed surface elastic constants ($C_{ijkl}$, GPa) for the bare Li slabs and the lowest $E_{form}$ systems (Figure 6) of each considered molecular gas after *NVT* MD equilibration. Bottom: changes in $C_{ijkl}$ ($\Delta C_{ijkl}$, GPa) with respect to the bare Li-slabs as a function of the system $E_{form}$.

## 4. Summary and conclusions

Using DFT geometry optimization and canonical *NVT* Molecular Dynamics, we have studied the adsorption of $N_2$, $O_2$, $CO_2$, $F_2$ and $SO_2$ on metal Li (100), (110) and (111) surfaces at 0 K and 300 K in the 0.25-1 ML coverage range. Structural, electronic and elastic characterization of the lowest energy systems indicates that:

i) All the considered gases interact exothermically with metal Li substrates. The reaction leads to profound rearrangement of the Li-slab accompanied by substantial Li→adsorbate charge transfer (≥ 1 electron/molecule). The Li-slab arrangement is accompanied by subsurface intercalation of the adsorbates and, in the case of $CO_2$, molecular condensation leading to acetylenediolate ($C_2O_2$) and subsurface oxide products.

ii) Depending on the dosed gas, different Li surface terminations can be energetically favored. Whereas $F_2$ treatment leaves the (100) surface energetically favored, $O_2$, $N_2$ and $CO_2$ dissociation leads to Li(110) being energetically favored. Contrary to the former cases, $SO_2$ treatment makes the Li(111) surface energetically favored.

iii) $SO_2$ is found to be the most effective gas in creating a thin insulating passivation layer. All the other gases lead to metallic adlayers.

iv) For the modeled 0.25-1 ML coverages, adlayer formation inevitably results in decrease of the Li-substrate work-function, suggesting detrimental enhancement of Li-reducing propensity and that, in line with experiments, molecular dosages larger than 1 ML are needed to chemically quench Li-substrate reactivity.

v) Apart from $N_2$, all the molecularly dissociated adlayers lead to substantial changes in the elastic properties of the slabs, with an overall tendency to elastic stiffness, or equivalently, reduced elastic compliance. Notably, $N_2$ dissociation into N-adatoms is found to yield the most elastically compliant adlayer. This suggests that small initial $N_2$ dosage could be beneficial for increased elasticity of the pristine passivation layer.

From a computational perspective, our study highlights also the following elements:

vi) Inclusion of van der Waals corrections in the simulations was explicitly tested and found to negligibly affect the dissociation of the considered molecular systems.

vii) Energy-screening of passivation layers based on geometry optimization of structures prepared from undissociated molecules turns out to be potentially misleading for Li substrates. Refinement of the screening via further optimization of *NVT* MD equilibrated snapshots invariably led to lower energy adlayer structures.

viii) The introduction of high vibrational frequencies, strongly exothermic slab formation energies and large Li→adsorbate charge transfer was found not to directly correlate with the computed adlayer elastic stiffening.

**Acknowledgments**

Support from the EU FP7 program SIRBATT (contract No. 608502) is gratefully acknowledged. SLK acknowledges Foundation of German Business for financial support. GT is supported by EPSRC-UK (EP/I004483/1). This work made use of the HPC Wales, N8 (EPSRC EP/K000225/1), ARCHER (via the UKCP Consortium, EP/K013610/1 and February 2014 RAP Allocation) High Performance Computing facilities as well as the computational resource bwUniCluster funded by the Ministry of Science, Research and the Arts Baden-Württemberg and the Universities of the State of Baden-Württemberg, Germany, within the framework program bwHPC.

**References**

[1] F. Croce, G.B. Appetecchi, L. Persi, B. Scrosati, Nanocomposite polymer electrolytes for lithium batteries, Nature 394 (1998) 456–458.

[2] J.-M. Tarascon, M. Armand, Issues and challenges facing rechargeable lithium batteries, Nature 414 (2001) 359–367.

[3] P.G. Bruce, S.A. Freunberger, L.J. Hardwick, J.-M. Tarascon, Li-O2 and Li-S batteries with high energy storage, Nat. Mater. 11 (2012) 19–29.

[4] X. Ji, K.T. Lee, L.F. Nazar, A highly ordered nanostructured carbon-sulphur cathode for lithium-sulphur batteries, Nat. Mater. 8 (2009) 500–506.

[5] H.-G. Jung, J. Hassoun, J.-B. Park, Y.-K. Sun, B. Scrosati, An improved high performance lithium-air battery, Nat. Chem. 4 (2012) 579–585.

[6] Z. Peng, S.A. Freunberger, Y. Chen, P.G. Bruce, A reversible and higher-rate Li-$O_2$ battery, Science 337 (2012) 563–566.

[7] Y. Chen, S.A. Freunberger, Z. Peng, O. Fontaine, P.G. Bruce, Charging a Li–O2 battery using a redox mediator, Nat. Chem. 5 (2013) 489-494.

[8] M.M.O. Thotiyl, S.A. Freunberger, Z. Peng, Y. Chen, Z. Liu, P.G. Bruce, A stable cathode for the aprotic Li–O2 battery, Nat. Mater. 12 (2013) 1050-1056.

[9] L. Johnson, C. Li, Z. Liu, Y. Chen, S.A. Freunberger, P.C. Ashok, B.B. Praveen, K. Dholakia, J.-M. Tarascon, P.G. Bruce, The role of $LiO_2$ solubility in $O_2$ reduction in aprotic solvents and its consequences for Li–$O_2$ batteries, Nat. Chem. 6 (2014) 1091-1099.

[10] D. R. Linde, CRC Handbook of Chemistry and Physics, 79th ed., Boca Raton, CRC Press, 1998.

[11] M. Winter, W.K. Appel, B. Evers, T. Hodal, K.-C. Möller, I. Schneider, M. Wachtler, M.R. Wagner, G.H. Wrodnigg, J.O. Besenhard, Studies on the anode/electrolyte interface in lithium ion batteries, Monatsh. Chem. 132 (2001) 473-486.

[12] D. Aurbach, A. Zaban, Y. Gofer, Y.E. Ely, I. Weissman, O. Chusid, O. Abramson, Recent studies of the lithium-liquid electrolyte interface electrochemical, morphological and spectral studies of a few important systems, J. Power Sources 54 (1995) 76–84.

[13] D. Aurbach, E. Zinigrad, H. Teller, P. Dan, Factors which limit the cycle life of rechargeable lithium (metal) batteries, J. Electrochem. Soc. 147 (2000) 1274–1279.

[14] P.C. Howlett, D.R. MacFarlane, A.F. Hollenkamp, High lithium metal cycling efficiency in a room-temperature ionic liquid, Electrochem. Solid-State Lett. 7 (2004) A97–A101.

[15] S.-K. Jeong, H.-Y. Seo, D.-H. Kim, H.-K. Han, J.-G. Kim, Y. B. Lee, Y. Iriyama, T. Abe, Z. Ogumi, Suppression of dendritic lithium formation by using concentrated electrolyte solutions, Electrochem. Commun. 10 (2008) 635–638.

[16] C. Fringant, A. Tranchant, R. Messina, Behavior of lithium-electrolyte interface during cycling in some ether-carbonate and carbonate mixtures, Electrochim. Acta 40 (1995) 513–523.

[17] T. Hirai, I. Yoshimatsu, J. Yamaki, Influence of electrolyte on lithium cycling efficiency with pressurized electrode stack, J. Electrochem. Soc. 141 (1994) 611–613.

[18] E. Eweka, J.R. Owen, A. Ritchie, Electrolytes and additives for high efficiency lithium cycling, J. Power Sources 65 (1997) 247–251.

[19] G.A. Umeda, E. Menke, M. Richard, K.L. Stamm, F. Wudl, B. Dunn, Protection of lithium metal surfaces using tetraethoxysilane, J. Mater. Chem. 21 (2011) 1593–1599.

[20] H. Ota, X. Wang, E. Yasukawa, Characterization of lithium electrode in lithium imides/ethylene carbonate, and cyclic ether electrolytes, J. Electrochem. Soc. 151 (2004) A427–A436.

[21] D. Aurbach, E. Zinigrad, Y. Cohen, H. Teller, A short review of failure mechanisms of lithium metal and lithiated graphite anodes in liquid electrolyte solution, Solid State Ion. 148 (2002) 405-416.

[22] J.B. Goodenough, Y. Kim, Challenges for rechargeable Li batteries, Chem. Mater. 22 (2010) 587-603.

[23] S. Kalnaus, A.S. Sabau, W.E. Tenhaeff, N.J. Dudney, C. Daniel, Design of composite polymer electrolytes for Li ion batteries based on mechanical stability criteria, J. Power Sources 201 (2012) 280-287.

[24] J.-i. Yamaki, S-i. Tobishima, K. Hayashi, K. Saito, Y. Nemoto, M. Arakawa, A consideration of the morphology of electrochemically deposited lithium in an organic electrolyte, J. Power Sources 74 (1998) 219–227.

[25] K.J. Harry, D.T. Hallinan, D.Y. Parkinson, A.A. MacDowell, N.P. Balsara, Detection of subsurface structures underneath dendrites formed on cycled lithium metal electrodes, Nat. Mater. 13 (2014) 69-73.

[26] T. Tatsuma, M. Taguchi, M. Iwaku, T. Sotomura, N. Oyama, Inhibition effects of polyacrylonitrile gel electrolytes on lithium dendrite formation, J. Electroanal. Chem. 472 (1999) 142-146.

[27] C. Monroe, J. Newman, Dendrite growth in lithium/polymer systems. A propagation model for liquid electrolytes under galvanostatic conditions, J. Electrochem. Soc. 150 (2003) A1377-A1384.

[28] M.Z. Mayers, J.W. Kaminski, T.F. Miller, Suppression of dendrite formation via pulse charging in rechargeable lithium metal batteries, J. Phys. Chem. C 116 (2012) 26214-26221.

[29] C. Monroe, J. Newman, The impact of elastic deformation on deposition kinetics at lithium/polymer interfaces, J. Electrochem. Soc. 152 (2005) A396-A404.

[30] M.S. Park, S.B. Ma, D.J. Lee, D. Im, S.-G. Doo, O. Yamamoto, A Highly Reversible Lithium Metal Anode, Sci. Rep. 4 (2014) 1-8.

[31] Y. Lu, Z. Tu, L.A. Archer, Stable lithium electrodeposition in liquid and nanoporous solid electrolytes, Nat. Mater. 13 (2014) 961–969.

[32] G. Zheng, S.W. Lee, Z. Liang, H.-W. Lee, K. Yan, H. Yao, H. Wang, W. Li, S. Chu, Y. Cui, Interconnected hollow carbon nanospheres for stable lithium metal anodes, Nat. Nanotechnol. 9 (2014) 618–623.

[33] R.O. Ritchie, The conflicts between strength and toughness, Nat. Mater. 10 (2011) 817–822.

[34] U.G.K. Wegst, H. Bai, E. Saiz, A.P. Tomsia, R.O. Ritchie, Bioinspired structural materials, Nat. Mater. 14 (2015) 23–36.

[35] D. Wolf, J.A. Jaszczak, Tailored elastic behavior of multilayers through controlled interface structure, J. Comput.-Aided Mol. Des. 1 (1993) 111-148.

[36] R.E. Miller, V.B. Shenoy, Size-dependent elastic properties of nanosized structural elements, Nanotechnology 11 (2000) 139-147.

[37] V.B. Shenoy, Atomistic calculations of elastic properties of metallic fcc crystal surfaces, Phys. Rev. B 71 (2005) 094104.

[38] V.B. Shenoy, Erratum: Atomistic calculations of elastic properties of metallic fcc crystal surfaces [Phys. Rev. B 71, 094104 (2005)], Phys. Rev. B 74 (2006) 149901(E).

[39] M. Wu, Z. Wen, Y. Liu, X. Wang, L. Huang, Electrochemical behaviors of a Li3N modified Li metal electrode in secondary lithium batteries, J. Power Sources 196 (2011) 8091–8097.

[40] M. Jäckle, A. Groß, Microscopic properties of lithium, sodium, and magnesium battery anode materials related to possible dendrite growth, J. Chem. Phys. 141 (2014) 174710.

[41] D. Gunceler, K. Letchworth-Weaver, R. Sundararaman, K.A. Schwarz, T.A. Arias, The importance of nonlinear fluid response in joint density-functional theory studies of battery systems, Modelling Simul. Mater. Sci. Eng. 21 (2013) 074005.

[42] Y. Ozhabes, D. Gunceler, T.A. Arias, Stability and surface diffusion at lithium-electrolyte interphases with connections to dendrite suppression, arXiv:1504.05799v1 (22$^{nd}$ April 2015).

[43] H. Valencia, M. Kohyama, S. Tanaka, H. Matsumoto, First-Principles Study of EMIM-FAFSA Molecule Adsorption on a Li(100) Surface as a Model for Li-Ion Battery Electrodes, J. Phys. Chem. C 116 (2012) 8493−8509.

[44] A. Budi, A. Basile, G. Opletal, A.F. Hollenkamp, A.S. Best, R.J. Rees, A.I. Bhatt, A.P. O'Mullane, S.P Russo, Study of the Initial Stage of Solid Electrolyte Interphase Formation upon Chemical Reaction of Lithium Metal and N‑Methyl‑N‑Propyl-Pyrrolidinium-Bis(Fluorosulfonyl)Imide, J. Phys. Chem. C 116 (2012) 19789−19797.

[45] S. Groh, M. Alam, Fracture behavior of lithium single crystal in the framework of (semi-)empirical force field derived from first-principles, Modelling Simul. Mater. Sci. Eng. 23 (2015) 045008.

[46] A. Aryanfar, D.J. Brooks, A.J. Colussi, B.V. Merinov, W.A. Goddard, M.R. Hoffmann, Thermal relaxation of lithium dendrites, Phys. Chem. Chem. Phys. 17 (2015) 8000-8005.

[47] T.J. Skotheim, C.J. Sheehan, Y.V. Mikhaylik, J.D. Affinito, Lithium anodes for electrochemical cells. Unites States Patent Application Publication, Pub No. US2014/0205912 A1, Pub. Date 24[th] July, 2014.

[48] D.Y. Wu, S. Meure, D. Solomon, Self-healing polymeric materials: A review of recent developments, Prog. Polym. Sci. 33 (2008) 479-522.

[49] T. Aida, E.W. Meijer, S.I. Stupp, Functional Supramolecular Polymers, Science 335 (2012) 813-817.

[50] N. Zhong, W. Post, Self-repair of structural and functional composites with intrinsically self-healing polymer matrices: A review, Composites, Part A 69 (2015) 226-239.

[51] T. Momma, H. Nara, S. Yamagami, C. Tatsumi, T. Osaka, Effect of the atmosphere on chemical composition and electrochemical properties of solid electrolyte interface on electrodeposited Li metal, J. Power Sources 196 (2011) 6483-6487.

[52] M. Nielsen, M.E. Björketun, M.H. Hansen, J. Rossmeisl, Towards first principles modeling of electrochemical electrode–electrolyte interfaces, Surf. Sci. 631 (2015) 2–7.

[53] N.G. Hörmann, M. Jäckle, F. Gossenberger, T. Roman, K. Forster-Tonigold, M. Naderian, S. Sakong, A. Groß, Some challenges in the first-principles modeling of structures and processes in electrochemical energy storage and transfer, J. Power Sources 275 (2015) 531–538.

[54] A.M. Souza, I. Rungger, C.D. Pemmaraju, U. Schwingenschloegl, S. Sanvito, Constrained-DFT method for accurate energy-level alignment of metal/molecule interfaces, Phys. Rev. B 88 (2013) 165112.

[55] G. Kresse, J. Furthmüller, Efficient iterative schemes for ab initio total-energy calculations using a plane-wave basis set, Phys. Rev. B 54 (1996) 11169.

[56] J.P. Perdew, K. Burke, M. Ernzerhof, Generalized Gradient Approximation Made Simple, Phys. Rev. Lett. 77 (1996) 3865-3868.

[57] P. Pulay, Convergence acceleration of iterative sequences. the case of SCF iteration, Chem. Phys. Lett. 73 (1980) 393-398.

[58] L. Verlet, Computer "Experiments" on Classical Fluids. I. Thermodynamical Properties of Lennard-Jones Molecules, Phys. Rev. 159 (1967) 98-103.

[59] H.J.C. Berendsen, J.P.M. Postma, W.F. van Gunsteren, A. DiNola, J.R. Haak, Molecular-Dynamics with Coupling to an External Bath, J. Chem. Phys. 81 (1984) 3684–3690.

[60] S.L. Kawahara, J. Lagoute, V. Repain, C. Chacon, Y. Girard, S. Rousset, A. Smogunov, C. Barreteau, Large Magnetoresistance through a Single Molecule due to a Spin-Split Hybridized Orbital, Nano Lett. 12 (2013) 4558-4563.

[61] T. Moorsom, M. Wheeler, T.M. Khan, F. Al Ma'Mari, C. Kinane, S. Langridge, D. Ciudad, A. Bedoya-Pinto, L. Hueso, G. Teobaldi, V.K. Lazarov, D. Gilks, G. Burnell, B.J. Hickey, O. Cespedes, Spin polarized electron transfer in ferromagnet/C60 interfaces, Phys. Rev. B 90 (2014) 125311.

[62] S. Grimme, Semiempirical GGA-Type Density Functional Constructed with a Long-Range Dispersion Correction, J. Comp. Chem. 27 (2006) 1787-1799.

[63] Y. Le Page, P. Saxe, Symmetry-general least-squares extraction of elastic data for strained materials from ab initio calculations of stress, Phys. Rev. B 65 (2002) 104104.

[64] X. Wu, D. Vanderbilt, D.R. Hamann, Systematic treatment of displacements, strains, and electric fields in density-functional perturbation theory, Phys. Rev. B 72 (2005) 035105.

[65] J.F. Nye, Physical Properties of Crystals, Oxford University Press, Oxford, 1985.

[66] G. Henkelman, A. Arnaldsson, H. Jónsson. A fast and robust algorithm for Bader decomposition of charge density, Comp. Mater. Sci. 36 (2006) 354-360.

[67] J. Wang, S.-Q. Wang, Surface energy and work function of fcc and bcc crystals: Density functional study, Surf. Sci. 630 (2014) 216–224.

[68] K. Doll, N.M. Harrison, V.R. Saunders, A density functional study of lithium bulk and surfaces, J. Phys.: Condens. Matter 11 (2009) 5007-5019.

[69] NOAA Earth System Research Laboratory, http://www.esrl.noaa.gov/gmd/

[70] P.E. Mason, F. Uhlig, V. Vaněk, T. Buttersack, S. Bauerecker, P. Jungwirth, Coulomb explosion during the early stages of the reaction of alkali metals with water, Nat. Chem. 7 (2015) 250–254.

[71] J.C. Rienstra-Kiracofe, G.S. Tschumper, H.F. Schaefer, Atomic and Molecular Electron Affinities: Photoelectron Experiments and Theoretical Computations, Chem. Rev. 102 (2002) 231-282.

[72] Isolated $N_2^-$ anion is unstable, accordingly experimental EA is not known (EA-1). Isolated $CO_2^-$ anion stable for 100 μs. See (EA-1) for further details.

[73] G. Zhuang, Y. Chen, P.N. Ross, The reaction of lithium with carbon dioxide studied by photoelectron spectroscopy, Surf. Sci. 418 (1998) 139–149.

[74] J. Wambach, G. Ordoefer, H.-J. Freund, H. Kuhlenbeck, M. Neumann, Influence of alkali co-adsorption on the adsorption and reaction of $CO_2$ on Pd(111), Surf. Sci. 209 (1989) 159-172.

[75] S. Wohlrab, D. Ehrlich, J. Wambach, H. Kuhlenbeck, H.-J. Freund, Promoter action of alkali in the activation of $CO_2$ on Pd(111): A HREELS case study, Surf. Sci. 220 (1989) 243-252.

[76] F.M. Hoffmann, M.D. Weisel, J. Paul, The activation of $CO_2$ by potassium-promoted Ru(001) I. FT-IRAS and TDMS study of oxalate and carbonate intermediates, Surf. Sci. 316 (1994) 277-293.

[77] Y. Shao, J. Paul, O. Axelsson, Identification of Intermediate States following $CO_2$ Adsorption on Alkali-Metal Surfaces, J. Phys. Chem. 97 (1993) 7652-7659.

[78] O. Axelsson, J. Paul, M.D. Weisel, F.M. Hoffmann, Reactive evaporation of potassium in $CO_2$ and the formation of bulk intermediates, J. Vac. Sci. Technol. A 12 (1994) 158-160.

[79] H.B. Michaelson, The work function of the elements and its periodicity, J. Appl. Phys. 48 (1977) 4729-4733.

[80] W.G. Hoover, Canonical dynamics: Equilibrium phase-space distributions, Phys. Rev. A 31 (1985) 1695–1697.

[81] A. Berces, T. Ziegler, Application of density functional theory to the calculation of force fields and vibrational frequencies of transition metal complexes. DENSITY FUNCTIONAL THEORY III. TOPICS IN CURRENT CHEMISTRY 182 (1996) 41-85.

[82] H. Burzlaff, H. Zimmermann, International Tables for Crystallography. Volume A, Section 10.1.2, Springer, Dordrecht, 2006.

**Appendix: Supplementary material**

**S1. Supplementary methods**

**S1.1. Surface Energies**

For the clean Li-slabs, surface energies ($E_{surface}$) were calculated as:

$$E_{surface} = \frac{E_{slab} - N_{Li} \cdot E_{Li-bulk}}{2A} \tag{S1}$$

where $E_{slab}$ is the energy of the optimized slab made up of $N_{Li}$ Li-atoms, $E_{Li-bulk}$ is the energy (per atom) of bulk Li(bcc) at the optimized lattice constant and $A$ is the slab surface area. The factor 2 accounts for the occurrence of two relaxed surfaces in the slab.

**S1.2. Surface energy and work function convergence with respect to slab thickness**

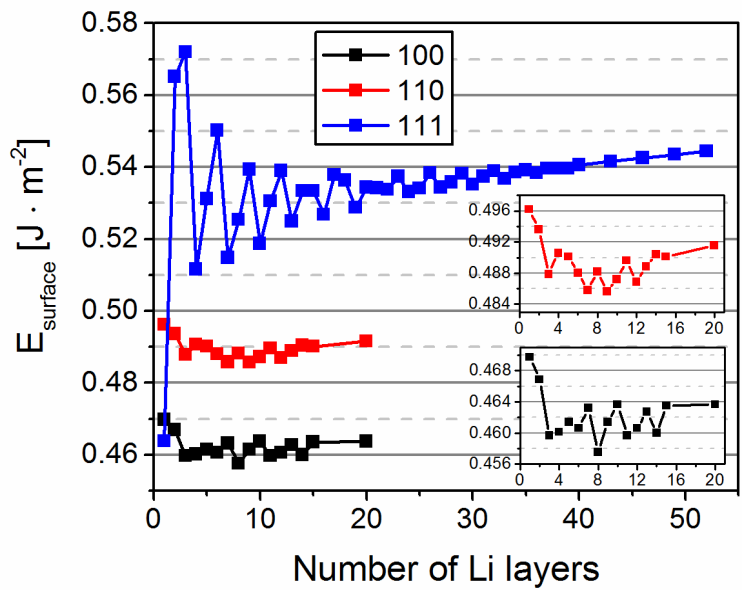

**Figure S1.** Computed convergence of the slab surface energy for Li(100), Li(110) and Li(111) as a function of the slab thickness. The insets reports a close up of the results for the Li(100) and Li(110) slabs.

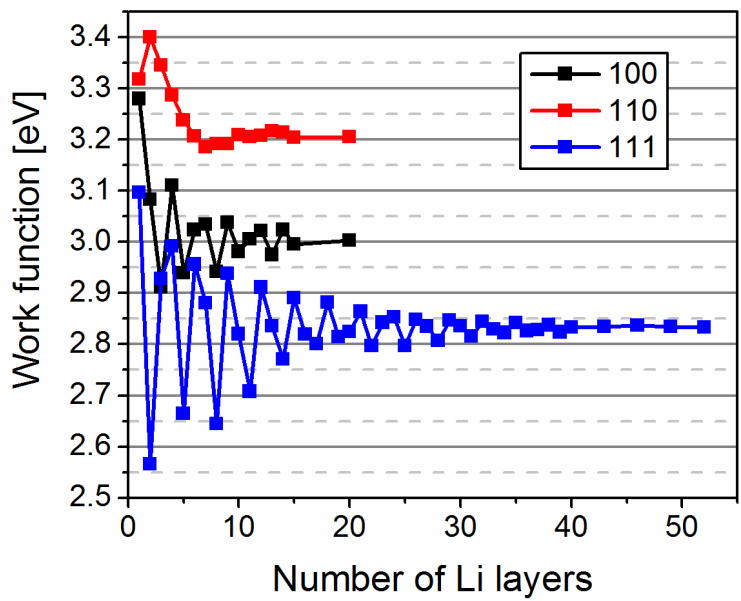

**Figure S2.** Computed convergence of the Li(100), Li(110) and Li(111) work function (W) as a function of the slab thickness.

### S1.3. Initial Geometries

All investigated gases ($N_2$, $O_2$, $CO_2$, $F_2$, $SO_2$) and single N-atoms ($N_{ad}$) were placed vertically and horizontally on the different adsorptions sites on both sides of the Li(100), Li(110) and Li(111) slabs shown in Figure S3 for several coverages in the 0.25–1 ML range. Table S1-S6 reports details of the considered initial adsorption geometries. We recall that each slab models contained four topmost Li-atoms per exposed surface (i.e. overall 8 topmost Li-atoms per slab).

For $SO_2$, the only considered non-linear molecule, vertical means that the O-S axis closer to the Li atoms was perpendicular to the slab. Horizontal $SO_2$ structures were prepared with either the S (h-S) or O-atom (h-O) closest to the Li slab.

Where applicable, the molecules were also rotated around the z-axis (= slab normal), so that the in-plane projection of the molecular axis was aligned with the different vectors shown in Figure S3. The initial closest distance between Li and adsorbate atoms was always 2.0 Å, except for $O_2$ (1.8 Å).

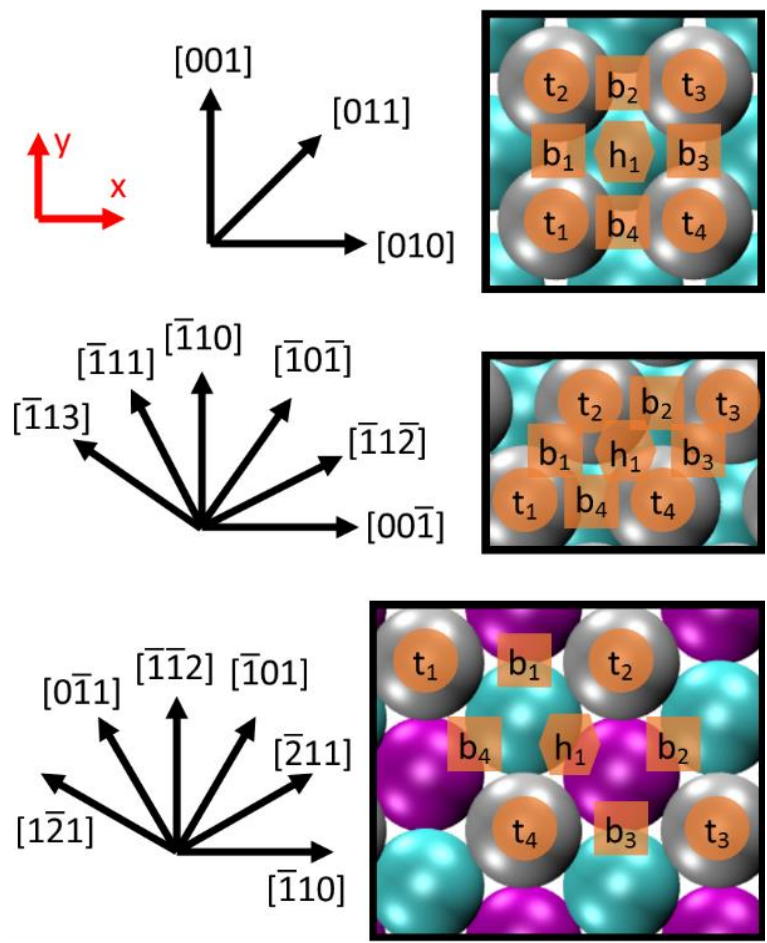

**Figure S3.** Scheme of the adopted labeling for the initial molecular adsorptions on from top to bottom: Li(100), Li(110) and Li(111).

## S2. Supplementary results

## S2.1. Energy screening of adsorption geometries

**Table S1.** The considered initial adsorption geometries and optimized $E_{form}$ for $F_2$ on Li(100), Li(110) and Li(111). The systems are sorted from lowest to highest $E_{form}$.

| Molecule | Facet | # molecules | ML | Initial adsorption sites | Vertical (v) or horizontal (h) adsorption | Orientation of the in-plane projection of the molecular axis | $E_{form}$ [eV] |
|---|---|---|---|---|---|---|---|
| $F_2$ | 100 | 8 | 1 | t1, t2, t3, t4 | h | [011] | -72.335 |
| | | 4 | 0.5 | t1, t4 | h | [011] | -42.059 |
| | | 4 | 0.5 | t1, t3 | h | [010] | -40.821 |
| | | 4 | 0.5 | t1, t4 | h | [010] | -40.815 |
| | | 4 | 0.5 | b2, b4 | v | | -39.094 |
| | | 4 | 0.5 | b2, b4 | h | [010] | -38.272 |
| | | 4 | 0.5 | t1, t3 | h | [011] | -34.523 |
| | | 8 | 1 | t1, t2, t3, t4 | h | [010] | -20.276 |
| | | 2 | 0.25 | b4 | h | [001] | -19.531 |
| | | 4 | 0.5 | t1, t3 | v | | -19.047 |
| | | 4 | 0.5 | b2, b4 | h | [001] | -18.513 |
| | | 2 | 0.25 | b4 | v | | -18.431 |
| | | 4 | 0.5 | t1, t4 | v | | -12.565 |
| | | 2 | 0.25 | b4 | h | [010] | -12.360 |
| | | 8 | 1 | t1, t2, t3, t4 | v | | -9.153 |
| | 110 | 4 | 0.5 | t1, t3 | h | [111] | -44.307 |
| | | 4 | 0.5 | t1, t3 | h | [111] | -44.289 |
| | | 4 | 0.5 | t1, t3 | h | [112] | -44.282 |
| | | 4 | 0.5 | b2, b4 | v | | -43.677 |
| | | 4 | 0.5 | t1, t3 | h | [001] | -42.637 |
| | | 4 | 0.5 | b1, b3 | h | [111] | -42.633 |
| | | 4 | 0.5 | b1, b3 | h | [001] | -42.606 |
| | | 4 | 0.5 | t1, t3 | v | | -41.890 |
| | | 4 | 0.5 | t1, t4 | h | [111] | -40.915 |
| | | 4 | 0.5 | t1, t4 | h | [112] | -40.897 |
| | | 4 | 0.5 | t1, t4 | h | [111] | -40.860 |
| | | 4 | 0.5 | b1, b3 | v | | -40.101 |
| | | 4 | 0.5 | b2, b4 | h | [001] | -39.495 |
| | | 4 | 0.5 | b2, b4 | h | [110] | -37.546 |
| | | 8 | 1 | t1, t2, t3, t4 | h | [111] | -22.691 |
| | | 8 | 1 | t1, t2, t3, t4 | h | [111] | -22.656 |
| | | 4 | 0.5 | t1, t4 | v | | -21.826 |
| | | 2 | 0.25 | b1 | h | [111] | -20.927 |
| | | 2 | 0.25 | b4 | h | [001] | -19.642 |
| | | 2 | 0.25 | b1 | h | [001] | -19.474 |
| | | 2 | 0.25 | b1 | v | | -18.970 |
| | | 2 | 0.25 | b4 | v | | -18.340 |
| | | 8 | 1 | t1, t2, t3, t4 | h | [112] | -17.446 |
| | | 8 | 1 | t1, t2, t3, t4 | h | [001] | -17.440 |
| | | 2 | 0.25 | b4 | h | [110] | -16.172 |
| | | 4 | 0.5 | t1, t4 | h | [001] | -14.384 |
| | | 8 | 1 | t1, t2, t3, t4 | v | | -8.231 |
| | 111 | 8 | 1 | t1, t2, t3, t4 | h | [011] | -85.390 |
| | | 8 | 1 | t1, t2, t3, t4 | h | [211] | -81.985 |
| | | 8 | 1 | t1, t2, t3, t4 | h | [101] | -81.979 |
| | | 8 | 1 | t1, t2, t3, t4 | h | [110] | -81.304 |
| | | 8 | 1 | t1, t2, t3, t4 | v | | -66.840 |
| | | 4 | 0.5 | b2, b4 | h | [111] | -41.107 |
| | | 4 | 0.5 | b1, b3 | h | [110] | -40.929 |
| | | 4 | 0.5 | t1, t3 | v | | -40.772 |
| | | 4 | 0.5 | t1, t3 | h | [011] | -40.632 |
| | | 4 | 0.5 | t1, t3 | h | [211] | -40.414 |
| | | 4 | 0.5 | b1, b3 | h | [001] | -40.372 |
| | | 4 | 0.5 | t1, t2 | h | [211] | -40.169 |
| | | 4 | 0.5 | t1, t3 | h | [101] | -40.129 |
| | | 4 | 0.5 | t1, t2 | h | [110] | -39.927 |
| | | 4 | 0.5 | t1, t3 | h | [110] | -39.345 |
| | | 4 | 0.5 | b2, b4 | h | [001] | -38.454 |
| | | 4 | 0.5 | b2, b4 | v | | -37.698 |
| | | 4 | 0.5 | t1, t2 | h | [101] | -37.323 |
| | | 4 | 0.5 | t1, t2 | h | [011] | -37.315 |
| | | 4 | 0.5 | b1, b3 | v | | -35.567 |
| | | 4 | 0.5 | t1, t2 | v | | -32.535 |
| | | 2 | 0.25 | b4 | h | [001] | -17.974 |
| | | 2 | 0.25 | b1 | h | [001] | -17.925 |
| | | 2 | 0.25 | b1 | h | [110] | -17.921 |
| | | 2 | 0.25 | b4 | h | [111] | -17.203 |
| | | 2 | 0.25 | b4 | v | | -16.622 |

**Table S2.** The considered initial adsorption geometries and optimized $E_{form}$ for $O_2$ on Li(100), Li(110) and Li(111). The systems are sorted from lowest to highest $E_{form}$.

| Molecule | Facet | # molecules | ML | Initial adsorption sites | Vertical (v) or horizontal (h) adsorption | Orientation of the in-plane projection of the molecular axis | $E_{form}$ [eV] |
|---|---|---|---|---|---|---|---|
| $O_2$ | 100 | 2 | 0.25 | b4 | h | [010] | -17.038 |
| | | 2 | 0.25 | b4 | h | [001] | -16.962 |
| | | 4 | 0.5 | b2, b4 | h | [010] | -16.386 |
| | | 4 | 0.5 | b2, b4 | v | | -16.036 |
| | | 4 | 0.5 | b2, b4 | h | [001] | -13.817 |
| | | 2 | 0.25 | b4 | v | | -6.723 |
| | | 4 | 0.5 | t1, t3 | h | [011] | -6.283 |
| | | 8 | 1 | t1, t2, t3, t4 | h | [010] | -5.694 |
| | | 8 | 1 | t1, t2, t3, t4 | h | [011] | -5.305 |
| | | 4 | 0.5 | t1, t3 | h | [010] | -5.261 |
| | | 8 | 1 | t1, t2, t3, t4 | v | | -4.485 |
| | | 4 | 0.5 | t1, t4 | h | [010] | -4.057 |
| | | 4 | 0.5 | t1, t4 | h | [011] | -4.014 |
| | | 4 | 0.5 | t1, t3 | v | | -2.101 |
| | | 4 | 0.5 | t1, t4 | v | | -1.577 |
| | 110 | 4 | 0.5 | b1, b3 | v | | -34.602 |
| | | 4 | 0.5 | t1, t3 | h | [111] | -34.325 |
| | | 4 | 0.5 | t1, t3 | h | [111] | -29.710 |
| | | 4 | 0.5 | b2, b4 | h | [001] | -29.707 |
| | | 4 | 0.5 | b1, b3 | h | [111] | -27.020 |
| | | 4 | 0.5 | b1, b3 | h | [001] | -26.619 |
| | | 2 | 0.25 | b1 | v | | -16.678 |
| | | 2 | 0.25 | b1 | h | [111] | -16.584 |
| | | 2 | 0.25 | b1 | h | [001] | -16.374 |
| | | 4 | 0.5 | t1, t3 | h | [001] | -16.302 |
| | | 4 | 0.5 | b2, b4 | v | | -16.282 |
| | | 4 | 0.5 | b2, b4 | h | [110] | -13.800 |
| | | 4 | 0.5 | t1, t3 | h | [112] | -13.066 |
| | | 4 | 0.5 | t1, t4 | h | [111] | -11.405 |
| | | 4 | 0.5 | t1, t4 | h | [112] | -11.391 |
| | | 4 | 0.5 | t1, t4 | h | [111] | -11.339 |
| | | 4 | 0.5 | t1, t3 | v | | -10.900 |
| | | 2 | 0.25 | b4 | h | [001] | -7.780 |
| | | 2 | 0.25 | b4 | v | | -7.212 |
| | | 4 | 0.5 | t1, t4 | h | [001] | -7.098 |
| | | 4 | 0.5 | t1, t4 | v | | -5.213 |
| | | 8 | 1 | t1, t2, t3, t4 | v | | -5.188 |
| | | 2 | 0.25 | b4 | h | [110] | -5.146 |

|  |  | 8 | 1 | t1, t2, t3, t4 | h | [001] | -3.852 |
|---|---|---|---|---|---|---|---|
|  |  | 8 | 1 | t1, t2, t3, t4 | h | [112] | -2.348 |
|  |  | 8 | 1 | t1, t2, t3, t4 | h | [111] | 4.849 |
|  |  | 8 | 1 | t1, t2, t3, t4 | h | [111] | 5.769 |
|  | 111 | 4 | 0.5 | t1, t2 | h | [211] | -37.570 |
|  |  | 4 | 0.5 | b2, b4 | h | [001] | -37.242 |
|  |  | 4 | 0.5 | t1, t3 | h | [110] | -37.070 |
|  |  | 4 | 0.5 | t1, t3 | v |  | -37.062 |
|  |  | 4 | 0.5 | t1, t3 | h | [101] | -36.476 |
|  |  | 4 | 0.5 | b2, b4 | h | [111] | -36.384 |
|  |  | 4 | 0.5 | b1, b3 | h | [110] | -36.274 |
|  |  | 4 | 0.5 | b1, b3 | v |  | -36.241 |
|  |  | 4 | 0.5 | t1, t3 | h | [211] | -35.972 |
|  |  | 4 | 0.5 | t1, t3 | h | [011] | -35.199 |
|  |  | 4 | 0.5 | t1, t2 | h | [011] | -34.206 |
|  |  | 4 | 0.5 | t1, t2 | h | [101] | -34.180 |
|  |  | 8 | 1 | t1, t2, t3, t4 | h | [211] | -33.620 |
|  |  | 8 | 1 | t1, t2, t3, t4 | h | [101] | -33.619 |
|  |  | 8 | 1 | t1, t2, t3, t4 | h | [110] | -33.480 |
|  |  | 4 | 0.5 | b2, b4 | v |  | -33.110 |
|  |  | 4 | 0.5 | b1, b3 | h | [001] | -32.648 |
|  |  | 8 | 1 | t1, t2, t3, t4 | h | [011] | -32.198 |
|  |  | 2 | 0.25 | b1 | v |  | -16.287 |
|  |  | 2 | 0.25 | b4 | h | [001] | -15.959 |
|  |  | 2 | 0.25 | b1 | h | [001] | -15.856 |
|  |  | 2 | 0.25 | b4 | h | [111] | -15.635 |
|  |  | 2 | 0.25 | b4 | v |  | -15.303 |
|  |  | 4 | 0.5 | t1, t2 | h | [110] | -13.958 |
|  |  | 2 | 0.25 | b1 | h | [110] | -4.725 |
|  |  | 8 | 1 | t1, t2, t3, t4 | v |  | -2.498 |
|  |  | 4 | 0.5 | t1, t2 | v |  | 0.519 |

**Table S3.** The considered initial adsorption geometries and optimized $E_{form}$ for $N_2$ on Li(100), Li(110) and Li(111). The systems are sorted from lowest to highest $E_{form}$.

| Molecule | Facet | # molecules | ML | Initial adsorption sites | Vertical (v) or horizontal (h) adsorption | Orientation of the in-plane projection of the molecular axis | $E_{form}$ [eV] |
|---|---|---|---|---|---|---|---|
| $N_2$ | 100 | 8 | 1 | t1, t2, t3, t4 | v | | -0.232 |
| | | 4 | 0.5 | t1, t4 | v | | 1.154 |
| | | 4 | 0.5 | t1, t3 | v | | 1.250 |
| | | 4 | 0.5 | b2, b4 | h | [010] | 1.473 |
| | | 4 | 0.5 | b2, b4 | v | | 1.636 |
| | | 2 | 0.25 | b4 | v | | 2.230 |
| | | 2 | 0.25 | b4 | h | [010] | 2.340 |
| | | 4 | 0.5 | t1, t4 | h | [010] | 2.369 |
| | | 4 | 0.5 | t1, t3 | h | [010] | 2.422 |
| | | 4 | 0.5 | t1, t3 | h | [011] | 2.469 |
| | | 4 | 0.5 | t1, t4 | h | [011] | 2.513 |
| | | 2 | 0.25 | b4 | h | [001] | 2.553 |
| | | 8 | 1 | t1, t2, t3, t4 | h | [011] | 3.052 |
| | | 4 | 0.5 | b2, b4 | h | [001] | 4.552 |
| | | 8 | 1 | t1, t2, t3, t4 | h | [010] | 6.413 |
| | 110 | 2 | 0.25 | b1 | h | [111] | -1.009 |
| | | 4 | 0.5 | b2, b4 | h | [001] | -0.736 |
| | | 4 | 0.5 | t1, t3 | h | [111] | -0.547 |
| | | 8 | 1 | t1, t2, t3, t4 | v | | -0.237 |
| | | 2 | 0.25 | b4 | h | [001] | -0.237 |
| | | 4 | 0.5 | t1, t4 | v | [001] | 0.416 |
| | | 4 | 0.5 | t1, t3 | v | | 0.455 |
| | | 4 | 0.5 | b2, b4 | v | | 0.650 |
| | | 4 | 0.5 | t1, t3 | h | [111] | 0.747 |
| | | 4 | 0.5 | b1, b3 | v | [001] | 0.752 |
| | | 4 | 0.5 | b2, b4 | h | [110] | 0.942 |
| | | 2 | 0.25 | b1 | v | | 1.285 |
| | | 2 | 0.25 | b4 | h | [110] | 1.375 |
| | | 4 | 0.5 | t1, t4 | h | [111] | 1.555 |
| | | 4 | 0.5 | t1, t4 | h | [111] | 1.601 |
| | | 2 | 0.25 | b4 | v | | 1.616 |
| | | 4 | 0.5 | t1, t4 | h | [112] | 1.633 |
| | | 4 | 0.5 | t1, t3 | h | [112] | 1.655 |
| | | 4 | 0.5 | t1, t3 | h | [001] | 1.691 |
| | | 2 | 0.25 | b1 | h | [001] | 1.824 |
| | | 4 | 0.5 | b1, b3 | h | [001] | 1.965 |
| | | 4 | 0.5 | b1, b3 | h | [111] | 1.971 |
| | | 4 | 0.5 | t1, t4 | h | [001] | 3.818 |
| | | 8 | 1 | t1, t2, t3, t4 | h | [112] | 7.626 |
| | | 8 | 1 | t1, t2, t3, t4 | h | [001] | 7.750 |
| | | 8 | 1 | t1, t2, t3, t4 | h | [111] | 9.961 |
| | | 8 | 1 | t1, t2, t3, t4 | h | [111] | 17.432 |
| | 111 | 4 | 0.5 | b2, b4 | h | [001] | 0.728 |
| | | 4 | 0.5 | b1, b3 | h | [110] | 0.775 |
| | | 8 | 1 | t1, t2, t3, t4 | v | | 2.203 |
| | | 2 | 0.25 | b4 | h | [001] | 2.630 |
| | | 2 | 0.25 | b1 | h | [110] | 2.767 |
| | | 4 | 0.5 | t1, t3 | v | | 3.433 |
| | | 4 | 0.5 | t1, t2 | v | | 3.434 |
| | | 8 | 1 | t1, t2, t3, t4 | h | [211] | 4.510 |
| | | 8 | 1 | t1, t2, t3, t4 | h | [110] | 4.514 |
| | | 8 | 1 | t1, t2, t3, t4 | h | [011] | 4.518 |
| | | 8 | 1 | t1, t2, t3, t4 | h | [101] | 4.533 |
| | | 4 | 0.5 | t1, t3 | h | [101] | 4.585 |
| | | 4 | 0.5 | t1, t2 | h | [101] | 4.587 |
| | | 4 | 0.5 | t1, t2 | h | [211] | 4.588 |
| | | 4 | 0.5 | t1, t2 | h | [011] | 4.592 |
| | | 4 | 0.5 | t1, t3 | h | [110] | 4.593 |
| | | 4 | 0.5 | t1, t3 | h | [011] | 4.594 |
| | | 4 | 0.5 | t1, t2 | h | [110] | 4.604 |
| | | 4 | 0.5 | t1, t3 | h | [211] | 4.613 |
| | | 4 | 0.5 | b2, b4 | h | [111] | 4.619 |
| | | 2 | 0.25 | b1 | h | [001] | 4.631 |
| | | 2 | 0.25 | b4 | h | [111] | 4.666 |
| | | 2 | 0.25 | b4 | v | | 4.686 |
| | | 2 | 0.25 | b1 | v | | 4.688 |
| | | 4 | 0.5 | b1, b3 | h | [001] | 4.711 |
| | | 4 | 0.5 | b1, b3 | v | | 4.760 |
| | | 4 | 0.5 | b2, b4 | v | | 4.766 |

**Table S4.** The considered initial adsorption geometries and optimized $E_{form}$ for $N_{ad}$ on Li(100), Li(110) and Li(111). The systems are sorted from lowest to highest $E_{form}$.

| Molecule | Facet | # molecules | ML | Initial adsorption sites | $E_{form}$ [eV] |
|---|---|---|---|---|---|
| $N_{ad}$ | 100 | 4 | 0.5 | b2, b4 | -2.029 |
| | | 2 | 0.25 | t1 | 0.207 |
| | | 2 | 0.25 | h1 | 0.231 |
| | | 4 | 0.5 | t1, t3 | 16.747 |
| | | 4 | 0.5 | t1, t4 | 18.072 |
| | | 8 | 1 | t1, t2, t3, t4 | 36.549 |
| | 110 | 8 | 1 | t1, t2, t3, t4 | -1.833 |
| | | 4 | 0.5 | t1, t4 | -1.140 |
| | | 2 | 0.25 | b1 | -0.903 |
| | | 2 | 0.25 | t1 | -0.824 |
| | | 2 | 0.25 | h1 | -0.744 |
| | | 4 | 0.5 | t1, t3 | -0.739 |
| | | 4 | 0.5 | b1, b3 | -0.710 |
| | | 2 | 0.25 | b4 | -0.667 |
| | | 4 | 0.5 | b2, b4 | -0.546 |
| | 111 | 4 | 0.5 | b2, b4 | -1.430 |
| | | 4 | 0.5 | t1, t3 | -0.194 |
| | | 2 | 0.25 | b1 | 2.117 |
| | | 2 | 0.25 | h1 | 2.151 |
| | | 2 | 0.25 | t1 | 12.511 |
| | | 4 | 0.5 | t1, t2 | 20.783 |
| | | 8 | 1 | t1, t2, t3, t4 | 38.291 |

**Table S5.** The considered initial adsorption geometries and optimized $E_{form}$ for $CO_2$ on Li(100), Li(110) and Li(111). The systems are sorted from lowest to highest $E_{form}$.

| Molecule | Facet | # molecules | ML | Initial adsorption sites | Vertical (v) or horizontal (h) adsorption | Orientation of the in-plane projection of the molecular axis | $E_{form}$ [eV] |
|---|---|---|---|---|---|---|---|
| $CO_2$ | 100 | 2 | 0.25 | h1 | h | [010] | -2.412 |
| | | 2 | 0.25 | h1 | h | [011] | -2.409 |
| | | 4 | 0.5 | b2, b4 | h | [010] | -0.844 |
| | | 4 | 0.5 | t1, t3 | h | [010] | 0.154 |
| | | 2 | 0.25 | b4 | h | [010] | 0.199 |
| | | 2 | 0.25 | t1 | h | [011] | 1.381 |
| | | 4 | 0.5 | t1, t4 | h | [010] | 1.421 |
| | | 2 | 0.25 | t1 | h | [010] | 1.423 |
| | | 4 | 0.5 | t1, t3 | v | [010] | 1.854 |
| | | 4 | 0.5 | t1, t4 | v | [010] | 1.989 |
| | | 2 | 0.25 | t1 | v | | 2.025 |
| | | 8 | 1 | t1, t2, t3, t4 | v | [010] | 2.048 |
| | | 2 | 0.25 | b4 | v | [010] | 2.244 |
| | | 2 | 0.25 | h1 | v | | 2.315 |
| | | 2 | 0.25 | b4 | h | [001] | 2.328 |
| | | 4 | 0.5 | b2, b4 | v | [010] | 2.399 |
| | | 4 | 0.5 | t1, t4 | h | [011] | 3.116 |
| | | 4 | 0.5 | t1, t3 | h | [011] | 3.814 |
| | | 8 | 1 | t1, t2, t3, t4 | h | [011] | 7.221 |
| | | 4 | 0.5 | b2, b4 | h | [001] | 19.188 |
| | | 8 | 1 | t1, t2, t3, t4 | h | [010] | 38.819 |
| | 110 | 4 | 0.5 | b2, b4 | h | [001] | -11.628 |
| | | 4 | 0.5 | t1, t3 | h | [111] | -6.470 |
| | | 4 | 0.5 | t1, t4 | h | [111] | -3.958 |
| | | 4 | 0.5 | t1, t4 | h | [112] | -3.958 |
| | | 4 | 0.5 | t1, t4 | h | [111] | -3.957 |
| | | 4 | 0.5 | t1, t3 | h | [111] | -3.792 |
| | | 2 | 0.25 | t1 | h | [111] | -3.503 |
| | | 2 | 0.25 | t1 | h | [111] | -3.495 |
| | | 2 | 0.25 | t1 | h | [112] | -3.479 |
| | | 2 | 0.25 | t1 | h | [001] | -3.326 |
| | | 2 | 0.25 | h1 | h | [111] | -3.321 |
| | | 2 | 0.25 | h1 | h | [111] | -3.277 |
| | | 2 | 0.25 | h1 | h | [112] | -3.277 |
| | | 2 | 0.25 | h1 | h | [001] | -3.276 |
| | | 4 | 0.5 | b2, b4 | h | [110] | -2.996 |
| | | 4 | 0.5 | t1, t3 | h | [112] | -1.963 |
| | | 2 | 0.25 | b4 | h | [001] | -1.855 |

|  |  | 2 | 0.25 | b1 | h | [001] | -1.584 |
|---|---|---|---|---|---|---|---|
|  |  | 4 | 0.5 | t1, t3 | h | [001] | -1.024 |
|  |  | 2 | 0.25 | b4 | h | [110] | -0.697 |
|  |  | 4 | 0.5 | b1, b3 | h | [111] | -0.154 |
|  |  | 4 | 0.5 | b1, b3 | h | [001] | -0.151 |
|  |  | 4 | 0.5 | b2, b4 | v |  | 1.361 |
|  |  | 2 | 0.25 | t1 | v |  | 1.400 |
|  |  | 4 | 0.5 | t1, t3 | v |  | 1.432 |
|  |  | 4 | 0.5 | t1, t4 | v |  | 1.473 |
|  |  | 2 | 0.25 | h1 | v |  | 1.577 |
|  |  | 2 | 0.25 | b1 | v |  | 1.589 |
|  |  | 2 | 0.25 | b4 | v |  | 1.685 |
|  |  | 4 | 0.5 | b1, b3 | v |  | 1.840 |
|  |  | 2 | 0.25 | b1 | h | [111] | 2.503 |
|  |  | 8 | 1 | t1, t2, t3, t4 | v |  | 3.511 |
|  |  | 4 | 0.5 | t1, t4 | h | [001] | 18.212 |
|  |  | 8 | 1 | t1, t2, t3, t4 | h | [112] | 44.541 |
|  |  | 8 | 1 | t1, t2, t3, t4 | h | [001] | 44.593 |
|  |  | 8 | 1 | t1, t2, t3, t4 | h | [111] | 49.430 |
|  |  | 8 | 1 | t1, t2, t3, t4 | h | [111] | 51.328 |
|  | 111 | 4 | 0.5 | b1, b3 | h | [110] | -6.747 |
|  |  | 4 | 0.5 | b2, b4 | h | [001] | -3.056 |
|  |  | 2 | 0.25 | h1 | h | [011] | -1.122 |
|  |  | 2 | 0.25 | b4 | h | [001] | -1.120 |
|  |  | 2 | 0.25 | h1 | h | [110] | -1.120 |
|  |  | 2 | 0.25 | b1 | h | [110] | -0.973 |
|  |  | 2 | 0.25 | h1 | h | [101] | -0.957 |
|  |  | 2 | 0.25 | t1 | h | [011] | -0.557 |
|  |  | 2 | 0.25 | t1 | h | [101] | -0.464 |
|  |  | 2 | 0.25 | t1 | h | [110] | -0.446 |
|  |  | 8 | 1 | t1, t2, t3, t4 | h | [211] | 2.974 |
|  |  | 8 | 1 | t1, t2, t3, t4 | h | [101] | 3.082 |
|  |  | 8 | 1 | t1, t2, t3, t4 | v |  | 3.167 |
|  |  | 2 | 0.25 | t1 | h | [211] | 3.574 |
|  |  | 4 | 0.5 | b1, b3 | v |  | 3.583 |
|  |  | 4 | 0.5 | b2, b4 | h | [111] | 3.597 |
|  |  | 4 | 0.5 | b2, b4 | v |  | 3.608 |
|  |  | 4 | 0.5 | t1, t3 | h | [101] | 3.703 |
|  |  | 4 | 0.5 | t1, t3 | h | [011] | 3.712 |
|  |  | 4 | 0.5 | t1, t3 | h | [211] | 3.736 |
|  |  | 4 | 0.5 | t1, t2 | h | [211] | 3.777 |
|  |  | 4 | 0.5 | t1, t2 | h | [101] | 3.783 |
|  |  | 4 | 0.5 | t1, t2 | v |  | 3.798 |
|  |  | 4 | 0.5 | t1, t3 | v |  | 3.798 |
|  |  | 2 | 0.25 | b1 | v |  | 4.102 |
|  |  | 2 | 0.25 | b4 | v |  | 4.117 |

|   |   | 2 | 0.25 | t1 | v |  | 4.183 |
|---|---|---|------|----|----|----|-------|
|   |   | 4 | 0.5 | b1, b3 | h | [001] | 4.479 |
|   |   | 2 | 0.25 | h1 | h | [211] | 4.526 |
|   |   | 2 | 0.25 | b1 | h | [001] | 4.526 |
|   |   | 2 | 0.25 | b4 | h | [111] | 4.566 |
|   |   | 2 | 0.25 | h1 | v |  | 4.617 |
|   |   | 4 | 0.5 | t1, t2 | h | [011] | 4.628 |
|   |   | 4 | 0.5 | t1, t3 | h | [110] | 4.646 |
|   |   | 8 | 1 | t1, t2, t3, t4 | h | [011] | 4.949 |
|   |   | 4 | 0.5 | t1, t2 | h | [110] | 5.168 |
|   |   | 8 | 1 | t1, t2, t3, t4 | h | [110] | 5.741 |

**Table S6.** The considered initial adsorption geometries and optimized $E_{form}$ for $SO_2$ on Li(100), Li(110) and Li(111). The systems are sorted from lowest to highest $E_{form}$.

| Molecule | Facet | # molecules | ML | Initial adsorption sites | Vertical (v) or horizontal (h) adsorption | Orientation of the in-plane projection of the molecular axis | $E_{form}$ [eV] |
|---|---|---|---|---|---|---|---|
| $SO_2$ | 100 | 4 | 0.5 | b2, b4 | v | [011] | -11.747 |
| | | 4 | 0.5 | t1, t3 | v | [011] | -10.499 |
| | | 4 | 0.5 | t1, t4 | v | [010] | -7.828 |
| | | 4 | 0.5 | t1, t4 | h-O | [010] | -7.748 |
| | | 4 | 0.5 | t1, t4 | h-O | [011] | -7.738 |
| | | 4 | 0.5 | t1, t3 | v | [010] | -7.731 |
| | | 4 | 0.5 | t1, t3 | h-O | [010] | -7.719 |
| | | 4 | 0.5 | b2, b4 | v | [010] | -7.023 |
| | | 4 | 0.5 | b2, b4 | h-O | [010] | -7.004 |
| | | 8 | 1 | t1, t2, t3, t4 | v | [011] | -5.947 |
| | | 8 | 1 | t1, t2, t3, t4 | v | [010] | -5.519 |
| | | 2 | 0.25 | b4 | v | [011] | -5.212 |
| | | 2 | 0.25 | t1 | v | [011] | -5.025 |
| | | 2 | 0.25 | h1 | h | [010] | -4.735 |
| | | 4 | 0.5 | t1, t3 | h-O | [011] | -4.599 |
| | | 2 | 0.25 | b4 | h-O | [001] | -4.369 |
| | | 2 | 0.25 | h1 | v | [010] | -4.276 |
| | | 4 | 0.5 | t1, t3 | h-S | [010] | -4.269 |
| | | 2 | 0.25 | h1 | v | [011] | -4.118 |
| | | 4 | 0.5 | t1, t4 | h-S | [011] | -4.043 |
| | | 4 | 0.5 | t1, t4 | h-S | [010] | -3.954 |
| | | 4 | 0.5 | b2, b4 | h-S | [001] | -3.726 |
| | | 4 | 0.5 | b2, b4 | h-S | [010] | -3.395 |
| | | 2 | 0.25 | h1 | h | [011] | -3.381 |
| | | 4 | 0.5 | t1, t4 | v | [011] | -2.255 |
| | | 2 | 0.25 | b4 | v | [010] | -2.218 |
| | | 2 | 0.25 | t1 | v | [010] | -2.140 |
| | | 2 | 0.25 | b4 | h-O | [010] | -1.879 |
| | | 2 | 0.25 | t1 | h | [010] | -0.884 |
| | | 2 | 0.25 | b4 | h-S | [010] | -0.819 |
| | | 4 | 0.5 | t1, t3 | h-S | [011] | -0.718 |
| | | 8 | 1 | t1, t2, t3, t4 | h-O | [011] | -0.492 |
| | | 2 | 0.25 | t1 | h | [011] | 0.763 |
| | | 2 | 0.25 | b4 | h-S | [001] | 0.990 |
| | | 4 | 0.5 | b2, b4 | h-O | [001] | 2.412 |
| | | 8 | 1 | t1, t2, t3, t4 | h-S | [011] | 7.481 |
| | | 8 | 1 | t1, t2, t3, t4 | h-O | [010] | 30.203 |
| | | 8 | 1 | t1, t2, t3, t4 | h-S | [010] | 37.790 |
| | 110 | 4 | 0.5 | t1, t3 | h-O | [111] | -21.043 |
| | | 4 | 0.5 | t1, t3 | h-O | [111] | -19.534 |
| | | 2 | 0.25 | h1 | h-S | [001] | -18.698 |
| | | 2 | 0.25 | h1 | h-S | [111] | -17.976 |
| | | 4 | 0.5 | t1, t3 | h-S | [111] | -14.603 |
| | | 4 | 0.5 | b2, b4 | h-S | [001] | -13.960 |
| | | 4 | 0.5 | b2, b4 | v | [001] | -12.899 |
| | | 4 | 0.5 | t1, t4 | v | [111] | -12.659 |
| | | 4 | 0.5 | b2, b4 | v | [111] | -12.217 |
| | | 4 | 0.5 | t1, t4 | v | [112] | -11.760 |
| | | 4 | 0.5 | t1, t3 | v | [111] | -11.719 |
| | | 2 | 0.25 | h1 | h-S | [111] | -11.715 |
| | | 4 | 0.5 | t1, t4 | v | [111] | -11.703 |
| | | 4 | 0.5 | b2, b4 | h-O | [001] | -11.676 |
| | | 4 | 0.5 | t1, t3 | v | [111] | -11.592 |
| | | 4 | 0.5 | t1, t4 | h-O | [111] | -11.529 |
| | | 4 | 0.5 | t1, t4 | h-O | [112] | -11.523 |
| | | 4 | 0.5 | t1, t4 | h-O | [111] | -11.454 |
| | | 4 | 0.5 | t1, t3 | v | [112] | -11.221 |
| | | 4 | 0.5 | b2, b4 | h-O | [111] | -10.653 |
| | | 4 | 0.5 | t1, t3 | h-S | [111] | -10.425 |
| | | 4 | 0.5 | b2, b4 | v | [110] | -10.044 |
| | | 4 | 0.5 | t1, t3 | v | [001] | -10.025 |
| | | 4 | 0.5 | t1, t3 | h-O | [112] | -8.733 |
| | | 4 | 0.5 | t1, t3 | h-O | [001] | -8.609 |
| | | 4 | 0.5 | t1, t4 | h-S | [111] | -8.544 |
| | | 4 | 0.5 | t1, t4 | h-S | [112] | -8.508 |
| | | 4 | 0.5 | t1, t4 | h-S | [111] | -8.501 |
| | | 8 | 1 | t1, t2, t3, t4 | v | [111] | -8.158 |
| | | 8 | 1 | t1, t2, t3, t4 | v | [112] | -8.143 |
| | | 4 | 0.5 | t1, t3 | h-S | [112] | -7.884 |
| | | 4 | 0.5 | b2, b4 | v | [001] | -7.417 |
| | | 4 | 0.5 | b2, b4 | h-S | [110] | -6.849 |
| | | 2 | 0.25 | t1 | v | [111] | -6.802 |
| | | 4 | 0.5 | b2, b4 | h-O | [110] | -6.794 |
| | | 2 | 0.25 | b4 | v | [001] | -6.694 |
| | | 4 | 0.5 | t1, t4 | v | [001] | -6.607 |
| | | 2 | 0.25 | h1 | v | [111] | -6.516 |
| | | 2 | 0.25 | h1 | v | [111] | -6.347 |
| | | 2 | 0.25 | t1 | v | [111] | -6.275 |
| | | 2 | 0.25 | t1 | v | [112] | -6.230 |
| | | 2 | 0.25 | b4 | v | [111] | -6.212 |
| | | 4 | 0.5 | b2, b4 | h-O | [001] | -6.095 |
| | | 2 | 0.25 | h1 | v | [001] | -6.073 |
| | | 2 | 0.25 | b4 | v | [001] | -5.638 |
| | | 2 | 0.25 | b4 | h-S | [111] | -5.613 |

|  |  | 2 | 0.25 | b4 | h-O | [111] | -5.613 |
|---|---|---|---|---|---|---|---|
|  |  | 2 | 0.25 | b4 | h-O | [001] | -5.601 |
|  |  | 2 | 0.25 | t1 | v | [001] | -5.529 |
|  |  | 2 | 0.25 | t1 | h-S | [112] | -5.520 |
|  |  | 2 | 0.25 | t1 | h-S | [111] | -5.475 |
|  |  | 2 | 0.25 | b4 | h-O | [001] | -5.423 |
|  |  | 4 | 0.5 | t1, t3 | h-S | [001] | -5.364 |
|  |  | 2 | 0.25 | t1 | h-S | [111] | -5.192 |
|  |  | 8 | 1 | t1, t2, t3, t4 | v | [001] | -4.841 |
|  |  | 2 | 0.25 | h1 | h-S | [112] | -4.823 |
|  |  | 2 | 0.25 | b4 | v | [110] | -4.804 |
|  |  | 2 | 0.25 | h1 | v | [112] | -4.601 |
|  |  | 2 | 0.25 | b4 | h-S | [001] | -4.405 |
|  |  | 2 | 0.25 | b4 | h-S | [001] | -4.091 |
|  |  | 2 | 0.25 | b4 | h-O | [110] | -4.008 |
|  |  | 2 | 0.25 | t1 | h-S | [001] | -2.836 |
|  |  | 8 | 1 | t1, t2, t3, t4 | v | [111] | -2.703 |
|  |  | 4 | 0.5 | b2, b4 | h-S | [111] | -2.434 |
|  |  | 2 | 0.25 | b4 | h-S | [110] | -1.453 |
|  |  | 4 | 0.5 | b2, b4 | h-S | [001] | -0.574 |
|  |  | 4 | 0.5 | t1, t4 | h-O | [001] | 8.890 |
|  |  | 4 | 0.5 | t1, t4 | h-S | [001] | 10.817 |
|  |  | 8 | 1 | t1, t2, t3, t4 | h-O | [112] | 32.973 |
|  |  | 8 | 1 | t1, t2, t3, t4 | h-O | [111] | 33.576 |
|  |  | 8 | 1 | t1, t2, t3, t4 | h-O | [001] | 33.964 |
|  |  | 8 | 1 | t1, t2, t3, t4 | h-S | [111] | 35.769 |
|  |  | 8 | 1 | t1, t2, t3, t4 | h-O | [111] | 36.252 |
|  |  | 8 | 1 | t1, t2, t3, t4 | h-S | [111] | 39.163 |
|  |  | 8 | 1 | t1, t2, t3, t4 | h-S | [001] | 41.730 |
|  |  | 8 | 1 | t1, t2, t3, t4 | h-S | [112] | 41.733 |
|  | 111 | 8 | 1 | t1, t2, t3, t4 | h-O | [101] | -28.912 |
|  |  | 8 | 1 | t1, t2, t3, t4 | h-O | [211] | -28.188 |
|  |  | 4 | 0.5 | t1, t3 | h-S | [101] | -25.274 |
|  |  | 8 | 1 | t1, t2, t3, t4 | h-O | [110] | -24.965 |
|  |  | 8 | 1 | t1, t2, t3, t4 | h-O | [011] | -24.587 |
|  |  | 2 | 0.25 | h1 | h-S | [101] | -17.288 |
|  |  | 4 | 0.5 | b1, b3 | v | [001] | -12.977 |
|  |  | 4 | 0.5 | b1, b3 | h-O | [001] | -12.804 |
|  |  | 4 | 0.5 | b1, b3 | h-O | [110] | -12.704 |
|  |  | 4 | 0.5 | t1, t3 | h-O | [110] | -12.672 |
|  |  | 4 | 0.5 | t1, t3 | v | [110] | -12.542 |
|  |  | 4 | 0.5 | b1, b3 | h-S | [110] | -12.455 |
|  |  | 4 | 0.5 | t1, t3 | v | [101] | -12.301 |
|  |  | 4 | 0.5 | t1, t3 | h-S | [011] | -12.271 |
|  |  | 4 | 0.5 | t1, t3 | h-O | [101] | -12.200 |
|  |  | 4 | 0.5 | t1, t2 | h-O | [101] | -12.196 |

| | | | | | | |
|---|---|---|---|---|---|---|
| | | 4 | 0.5 | t1, t2 | h-O | [011] | -12.189 |
| | | 4 | 0.5 | b2, b4 | h-O | [111] | -11.990 |
| | | 4 | 0.5 | t1, t3 | h-O | [011] | -11.820 |
| | | 4 | 0.5 | t1, t3 | h-O | [211] | -11.759 |
| | | 4 | 0.5 | b2, b4 | v | [111] | -11.731 |
| | | 4 | 0.5 | b2, b4 | h-S | [111] | -11.724 |
| | | 4 | 0.5 | b2, b4 | v | [001] | -11.712 |
| | | 4 | 0.5 | t1, t2 | v | [011] | -11.675 |
| | | 4 | 0.5 | t1, t2 | v | [101] | -11.672 |
| | | 4 | 0.5 | b1, b3 | h-S | [001] | -11.281 |
| | | 4 | 0.5 | t1, t2 | h-O | [110] | -11.247 |
| | | 4 | 0.5 | b1, b3 | v | [110] | -11.218 |
| | | 4 | 0.5 | t1, t2 | h-O | [211] | -10.721 |
| | | 4 | 0.5 | b2, b4 | h-S | [001] | -9.524 |
| | | 4 | 0.5 | b2, b4 | h-O | [001] | -7.960 |
| | | 4 | 0.5 | t1, t3 | v | [211] | -7.484 |
| | | 8 | 1 | t1, t2, t3, t4 | v | [211] | -4.731 |
| | | 8 | 1 | t1, t2, t3, t4 | v | [110] | -4.629 |
| | | 8 | 1 | t1, t2, t3, t4 | v | [101] | -4.626 |
| | | 8 | 1 | t1, t2, t3, t4 | v | [011] | -4.624 |
| | | 2 | 0.25 | b1 | v | [001] | -4.351 |
| | | 2 | 0.25 | h1 | h-S | [110] | -4.192 |
| | | 2 | 0.25 | h1 | v | [110] | -3.928 |
| | | 2 | 0.25 | h1 | v | [101] | -3.923 |
| | | 2 | 0.25 | b4 | h-O | [001] | -3.874 |
| | | 2 | 0.25 | h1 | v | [211] | -3.838 |
| | | 2 | 0.25 | b4 | v | [001] | -3.801 |
| | | 2 | 0.25 | t1 | v | [101] | -3.750 |
| | | 2 | 0.25 | b1 | v | [110] | -3.691 |
| | | 2 | 0.25 | b4 | h-O | [111] | -3.476 |
| | | 2 | 0.25 | h1 | v | [011] | -3.451 |
| | | 2 | 0.25 | b4 | v | [111] | -3.443 |
| | | 2 | 0.25 | t1 | v | [011] | -3.430 |
| | | 2 | 0.25 | b1 | h-O | [001] | -3.333 |
| | | 2 | 0.25 | b1 | h-O | [110] | -3.293 |
| | | 2 | 0.25 | b1 | h-S | [001] | -3.146 |
| | | 2 | 0.25 | h1 | h-S | [211] | -2.145 |
| | | 2 | 0.25 | b4 | h-S | [111] | -2.127 |
| | | 2 | 0.25 | b4 | h-S | [001] | -1.850 |
| | | 2 | 0.25 | h1 | h-S | [011] | -1.843 |
| | | 2 | 0.25 | b1 | h-S | [110] | -1.016 |
| | | 4 | 0.5 | t1, t2 | v | [211] | -0.200 |
| | | 4 | 0.5 | t1, t3 | v | [011] | -0.154 |
| | | 4 | 0.5 | t1, t2 | v | [110] | -0.119 |
| | | 2 | 0.25 | t1 | v | [110] | 0.035 |
| | | 2 | 0.25 | t1 | v | [211] | 0.048 |

|  |  |  |  |  |  |  |
|---|---|---|---|---|---|---|
|  |  | 4 | 0.5 | t1, t3 | h-S | [211] | 2.671 |
|  |  | 4 | 0.5 | t1, t2 | h-S | [101] | 2.702 |
|  |  | 4 | 0.5 | t1, t2 | h-S | [011] | 2.702 |
|  |  | 8 | 1 | t1, t2, t3, t4 | h-S | [211] | 2.708 |
|  |  | 4 | 0.5 | t1, t3 | h-S | [110] | 2.712 |
|  |  | 4 | 0.5 | t1, t2 | h-S | [211] | 2.719 |
|  |  | 8 | 1 | t1, t2, t3, t4 | h-S | [011] | 2.720 |
|  |  | 8 | 1 | t1, t2, t3, t4 | h-S | [101] | 2.729 |
|  |  | 2 | 0.25 | t1 | h-S | [110] | 3.348 |
|  |  | 2 | 0.25 | t1 | h-S | [211] | 3.350 |
|  |  | 2 | 0.25 | t1 | h-S | [101] | 3.353 |
|  |  | 2 | 0.25 | t1 | h-S | [011] | 3.363 |
|  |  | 4 | 0.5 | t1, t2 | h-S | [110] | 4.130 |
|  |  | 8 | 1 | t1, t2, t3, t4 | h-S | [110] | 4.572 |

**Table S7.** The lowest $E_{form}$ (eV) systems and corresponding work function (W) for all the considered gas adsorbed on Li(100), Li(110) and Li(111). Same data as in Figs. 2 and 5.

| Adsorbate | Facet | $N_{mol}$ | $E_{form}$ (eV) (before MD) | $E_{form}$ (eV) (after MD) | W (eV) (before MD) | W (eV) (after MD) |
|---|---|---|---|---|---|---|
| $F_2$ | 100 | 8 | -72.335 | -89.873 | 2.548 | 1.850 |
| | 110 | 4 | -44.307 | -44.502 | 2.226 | 2.161 |
| | 111 | 8 | -85.390 | -86.492 | 1.768 | 1.934 |
| $O_2$ | 100 | 2 | -17.038 | -17.152 | 2.772 | 2.773 |
| | 110 | 4 | -34.602 | -38.141 | 2.074 | 2.473 |
| | 111 | 4 | -37.570 | -37.565 | 1.937 | 1.957 |
| $N_2$ | 100 | 8 | -0.236 | 0.531 | 3.835 | 3.613 |
| | 110 | 2 | -1.009 | -1.411 | 2.510 | 2.731 |
| | 111 | 4 | 0.729 | -1.775 | 2.557 | 2.413 |
| $N_{ad}$ | 100 | 4 | -2.029 | -2.781 | 2.252 | 2.139 |
| | 110 | 4 | -3.145 | -3.165 | 2.496 | 2.499 |
| | 111 | 4 | -1.430 | -1.629 | 2.321 | 2.267 |
| $CO_2$ | 100 | 4 | -0.844 | -7.815 | 4.952 | 2.293 |
| | 110 | 4 | -11.629 | -13.240 | 2.841 | 2.703 |
| | 111 | 4 | -6.747 | -6.554 | 2.283 | 2.268 |
| $SO_2$ | 100 | 4 | -11.747 | -12.797 | 2.497 | 1.903 |
| | 110 | 4 | -21.059 | -21.663 | 3.867 | 3.075 |
| | 111 | 8 | -28.912 | -29.152 | 3.119 | 2.610 |

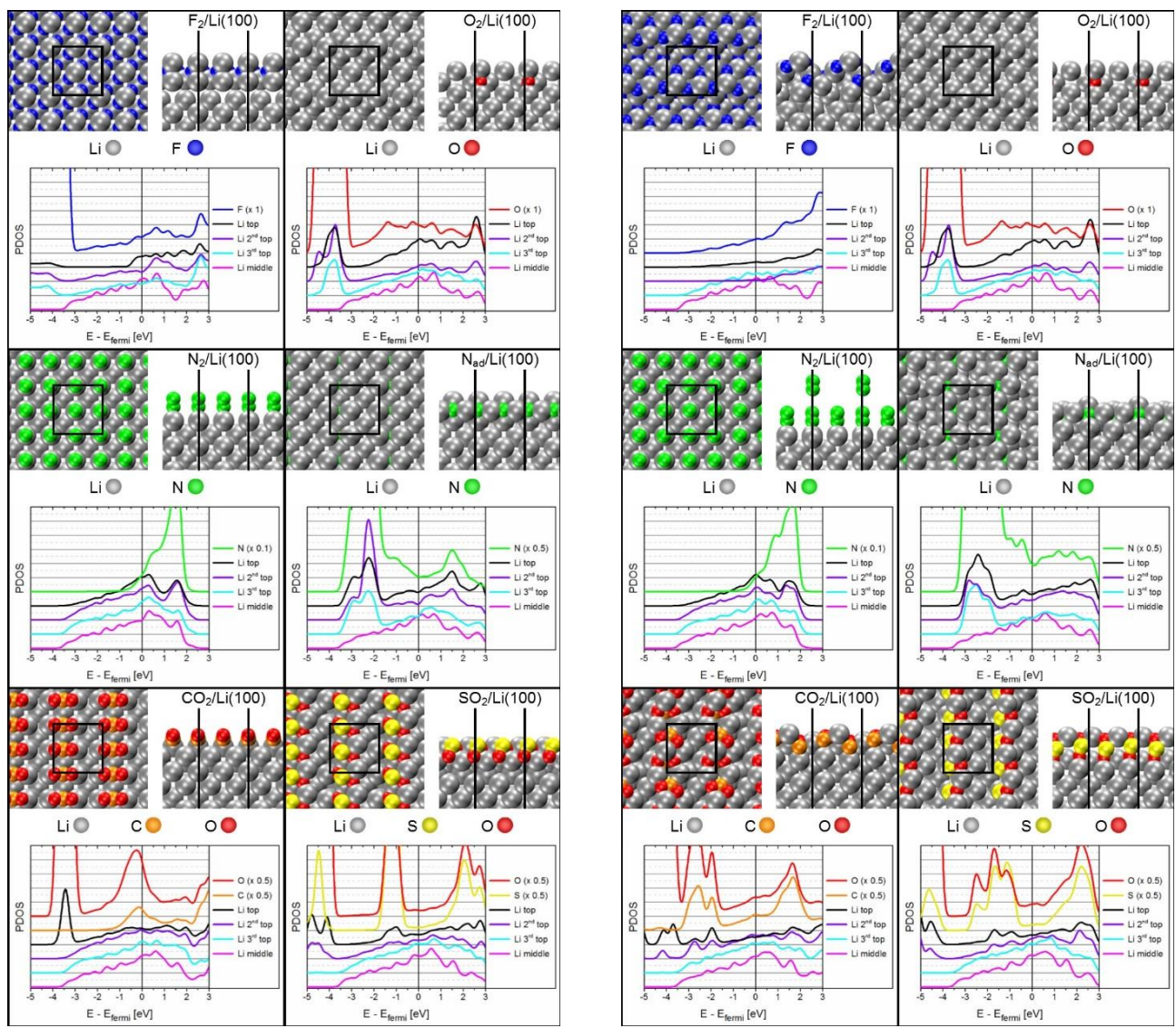

**Figure S4.** Optimized geometry and layer-resolved atom-projected Density of States (PDOS) for the lowest $E_{form}$ systems (Table S7) on **Li(100)** before (left) and after (right) NVT MD equilibration.

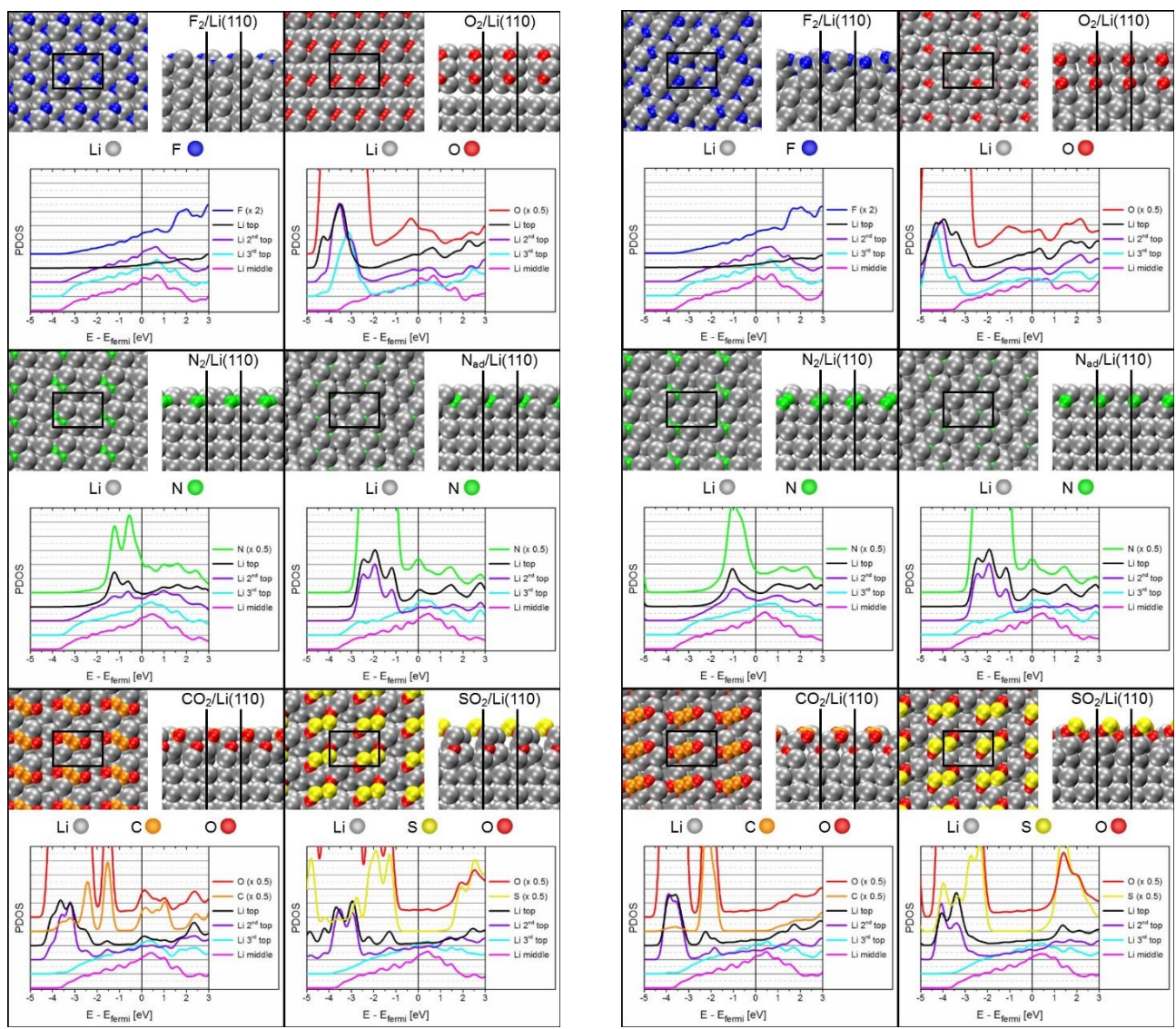

**Figure S5.** Optimized geometry and layer-resolved atom-projected Density of States (PDOS) for the lowest $E_{form}$ systems (Table S7) on **Li(110)** before (left) and after (right) NVT MD equilibration.

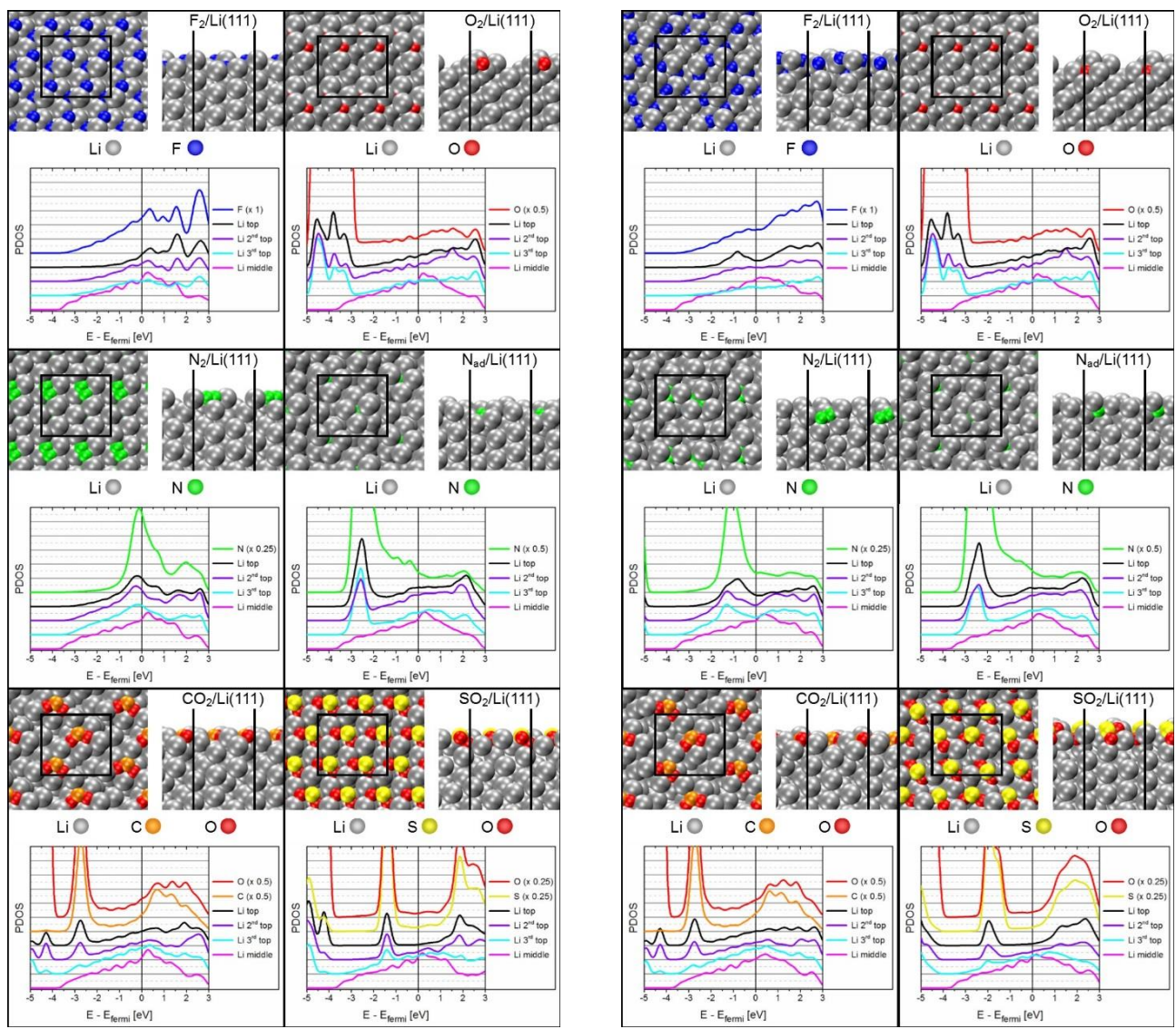

**Figure S6.** Optimized geometry and layer-resolved atom-projected Density of States (PDOS) for the lowest $E_{form}$ systems (Table S7) on **Li(111)** before (left) and after (right) NVT MD equilibration.

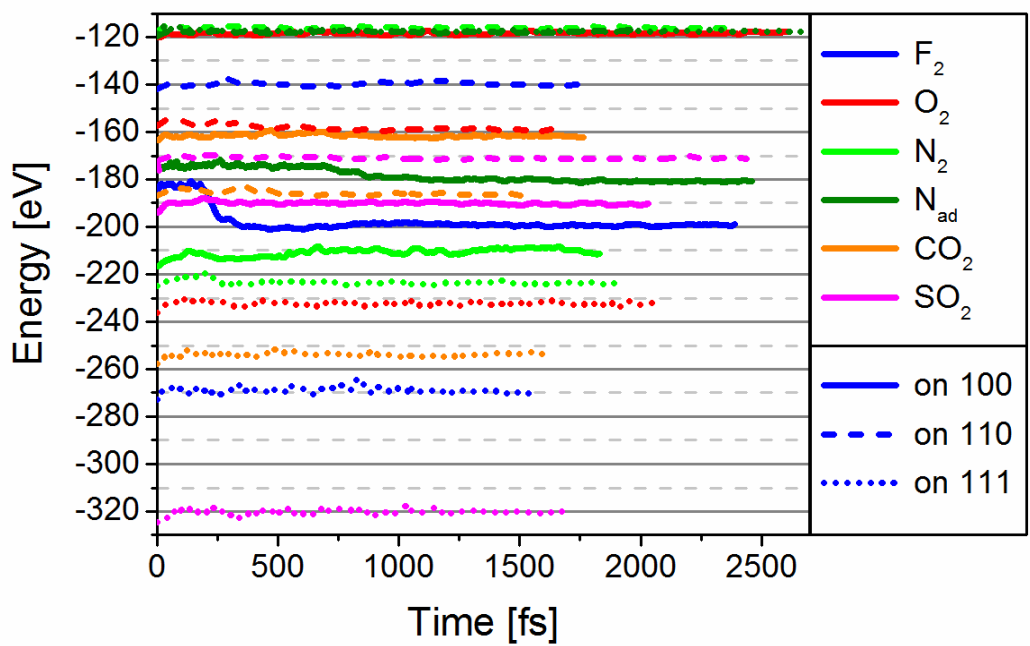

**Figure S7.** DFT energy as a function of time for the lowest $E_{form}$ optimized systems on Li(100), Li(110) and Li(111), see also Table S7.

**Table S8.** Selected shortest inter-atomic distances (Å) for the lowest $E_{form}$ systems **before** NVT MD equilibration (Table S7).

| System | |
|---|---|
| $F_2$/Li(111) | **Li-F:** 1.76, 1.80, 1.85, 3.11, 3.49, 3.54 <br> **F-F:** 2.88 |
| $O_2$/Li(111) | **Li-O:** 1.78, 1.80, 1.85, 1.88, 1.94, 3.50 <br> **O-O:** 2.94 |
| $N_2$/Li(110) | **Li-N:** 1.82, 1.89, 1.97, 2.04, 2.61, 3.10 <br> **N-N:** 1.34 |
| $N_{ad}$/Li(110) | **Li-N:** 1.82, 1.87, 1.89, 1.95, 1.98, 2.08, 3.99 <br> **N-N:** 3.21 |
| $CO_2$/Li(110) | **Li-C:** 2.10, 2.22, 2.36, 2.97, 3.06, 3.75 <br> **Li-O ($C_2O_2$):** 2.10, 2.22, 2.36, 2.97, 3.06, 3.75 <br> **Li-O (O-atom):** 1.76, 1.78, 1.83, 1.86, 2.02, 3.02 <br> **C-C:** 1.32 <br> **C-O ($C_2O_2$):** 1.31 <br> **C-O ($C_2O_2$-O):** 3.18 <br> **O-O ($C_2O_2$):** 3.22 <br> **O-O ($C_2O_2$-O):** 2.94 <br> **O-O (O-O):** 2.92 |
| $SO_2$/Li(111) | **Li-S:** 2.63, 2.88, 2.88, 2.97, 3.05, 3.10, 3.27 <br> **Li-O:** 1.85, 1.87, 3.04, 3.25, 3.54, 3.94 <br> **S-O:** 1.60, 1.66 <br> **O-O:** 2.57 |

**Table S9.** Selected shortest inter-atomic distances (Å) for the lowest $E_{form}$ systems **after** NVT MD equilibration (Table S7).

| System | |
|---|---|
| $F_2$/Li(100) | **Li-F:** 1.89, 1.92, 1.93, 1.94, 3.17, 3.52 <br> **F-F:** 2.76 |
| $O_2$/Li(110) | **Li-O:** 1.76, 1.85, 1.86, 1.91, 1.96, 1.96 <br> **O-O:** 2.94 |
| $N_2$/Li(111) | **Li-N:** 1.97, 2.02, 2.03, 2.04, 2.07, 2.08, 2.88 <br> **N-N:** 1.44 |
| $N_{ad}$/Li(110) | **Li-N:** 1.82, 1.87, 1.89, 1.94, 1.98, 2.09, 3.96 <br> **N-N:** 3.21 |
| $CO_2$/Li(110) | **Li-C:** 2.04, 2.41, 2.75, 3.20, 3.74, 3.97 <br> **Li-O ($C_2O_2$):** 1.97, 1.98, 2.00, 3.15, 3.21, 3.57, 4.11 <br> **Li-O (O-atom):** 1.82, 1.82, 1.83, 1.87, 1.92, 3.59 <br> **C-C:** 1.25 <br> **C-O ($C_2O_2$):** 1.33 <br> **C-O ($C_2O_2$-O):** 3.13 <br> **O-O ($C_2O_2$):** 3.29 <br> **O-O ($C_2O_2$-O):** 3.14 <br> **O-O (O-O):** 3.52 |
| $SO_2$/Li(111) | **Li-S:** 2.79, 2.88, 2.95, 2.96, 3.00, 3.04, 3.20 <br> **Li-O:** 1.92, 1.93, 1.99, 2.89, 3.47, 3.69 <br> **S-O:** 1.64, 1.65 <br> **O-O:** 2.56 |